\documentclass[a4paper,12pt]{article}
\usepackage{amsmath, amssymb, amsthm}
\usepackage{fullpage}
\usepackage{hyperref}
\usepackage{graphicx}
\usepackage{parskip}
\usepackage{color}
\usepackage{appendix}
\usepackage{subcaption}
\usepackage{multirow}
\usepackage[labelformat=simple]{subcaption}

\numberwithin{equation}{section}

\begin{document}
\title{Cabling in the Skyrme-Faddeev model}
\author{Paul Jennings\\[10pt]
{\em \normalsize Department of Mathematical Sciences, }\\{\em \normalsize Durham University, Durham, DH1 3LE, U.K.}\\[10pt]
{\normalsize paul.jennings@durham.ac.uk}}
\date{December 2014}

\maketitle
\vspace{15pt}
\begin{abstract}
\noindent
The Skyrme-Faddeev model is a three-dimensional non-linear field theory that has topological soliton solutions, called hopfions, which are novel string-like solutions taking the form of knots and links. Solutions found thus far take the form of torus knots and links of these, however torus knots form only a small family of known knots. It is an open question whether any non-torus knot hopfions exist. In this paper we present a construction of knotted fields with the form of cable knots to which an energy minimisation scheme can be applied. We find the first known hopfions which do not have the form of torus knots, but instead take the form of cable and hyperbolic knots.
\end{abstract}

\pagebreak

\section{Introduction}

The Skyrme-Faddeev model \cite{faddeev1975,faddeev1997} is a $(3+1)$-dimensional  modified $O(3)$ sigma-model that includes a term quartic in derivatives to provide a preferred scale. This model exhibits topological solitons, called hopfions, which have the form of stable closed strings which form knots and links. The Skyrme-Faddeev model is closely related to the other Skyrme models. It can be thought of as a restriction of the target space in the Skyrme model \cite{skyrme1961}, or as a string of baby Skyrmions \cite{piette1995}, with each baby Skyrmion living in the transverse plane to the centre curve of the hopfion. The Skyrme-Faddeev model has links to low-energy QCD, with it having been suggested that this model may describe glueballs \cite{faddeev1998}, and applications have also been suggested in condensed matter systems \cite{kawaguchi2008}.

While no exact analytic solutions to the Skyrme-Faddeev model are known, previous studies have found numerical solutions for hopfions with topological charges of up to sixteen \cite{faddeev1997, gladikowski1997, battye1998, battye1999, hietarinta1999, hietarinta2000, ward2000, sutcliffe2007}. In recent studies a rational map approximation was implemented to construct fields with a desired structure and were of good approximation to solutions. The energy landscape has many local minima each with large basins of attraction that makes energy relaxation techniques more complicated. For higher charges there are more local minima, and so the benefits of a wide variety of initial conditions becomes paramount if one wishes to understand the energy minima for higher topological charge.

All solutions known to date are either torus knots or links formed from torus knot constituents and previous studies have left it open as to whether hopfions taking the form of non-torus knots in the Skyrme-Faddeev model exist. It is known  \cite{thurston1986,adams2004} that all knots can be classified into one of three groups: torus knots, satellite knots (of which cable knots are a subclass) and hyperbolic knots. In this paper we begin by giving details of the model we shall be studying, and discuss how fields with the desired topological structure can be generated as in \cite{sutcliffe2007} in Section \ref{sec::model}, before explaining how we may construct knotted fields with the form of cable knots in Section \ref{sec::cable}. In Section \ref{sec::results} we then present the results of applying an energy minimisation algorithm to the initial field configurations constructed for solutions for a variety of topological charges.

\section{The Skyrme-Faddeev model}\label{sec::model}

The Skyrme-Faddeev model involves a map $\boldsymbol{\phi}:\mathbb{R}^3\to S^2$, which we parametrise as a three-component unit vector. Since in this current paper we are considering only static fields, we may define the Skyrme-Faddeev model via the static energy functional of this field
\begin{equation}\label{eq::energy}
E=\frac{1}{32\pi^2\sqrt{2}}\int\partial_i\boldsymbol{\phi}\cdot \partial_i\boldsymbol{\phi}+\frac{1}{2}(\partial_i\boldsymbol{\phi}\times \partial_j\boldsymbol{\phi})\cdot(\partial_i\boldsymbol{\phi}\times \partial_j\boldsymbol{\phi})\,d^3x.
\end{equation}
For the field to have a finite energy it must tend towards a constant vector at spatial infinity, which we choose to be $\lim_{|\mathbf{x}|\to\infty}\boldsymbol{\phi}(\mathbf{x})=\boldsymbol{\phi}_\infty=(0,0,1)$ without loss of generality. This identification leads to a one-point compactification of physical space so the field is a map $\boldsymbol{\phi}:S^3 \to S^2$ and so has related homotopy group $\pi_3(S^2)=\mathbb{Z}$. Since this is a map between spaces of different dimensions, the topological charge can not simply be the degree of the map, as is the case for other Skyrme theories. Instead we find that the topological charge is given by the Hopf invariant. Let $f$ be the pullback under $\boldsymbol{\phi}$ of the area two-form on the target $S^2$. Then the triviality of the second cohomology group of $S^3$ means that $f$ must be exact, say $f=da$, and the Hopf charge is given by
\begin{equation}
Q=\frac{1}{4\pi^2}\int_{S^3}f\wedge a.
\end{equation}
The Hopf charge can not be written as a density which is local in terms of $\boldsymbol{\phi}$, and so it is useful to think of it in a more intuitive geometric way. The preimage of a point of $S^2$ is generically (a collection of) closed curves. The Hopf charge is then given by the linking number of the preimage of two distinct points of the target $S^2$ \cite{bott1995}. For this an orientation of each preimage curve is needed, which follows from the behaviour of the field about this preimage, as detailed by Hietarinta and Salo \cite{hietarinta1999,hietarinta2000}. An orientation follows from the direction of the winding of the field about the location of the soliton corresponding to a direction in the $(\phi_1,\phi_2)$ components of target space for a fixed value of $\phi_3$. In this study we restrict to solutions with positive Hopf charge, since solutions with a negative charge can be obtained by a spatial reflection, which changes the sign of the charge. It will also be useful to note that one can consider the Hopf charge as being composed of the crossing number of one curve plus the winding number of the second curve about this. However these individually are not topologically conserved quantities.

It is known that the energy functional \eqref{eq::energy} obeys the Vakulenko-Kapitanskii bound \cite{vakulenko1979}
\begin{equation}
E\ge c\,|Q|^{3/4},
\end{equation}
where the sublinear growth of the energy is due to the use of complicated Sobolev inequalities. This can be understood physically as a consequence of the creation of additional charge through the knotting and linking of solitons. The inequality has been proven for $c=(3/16)^{3/8}\approx0.534$ \cite{kundu1982}, although it is believed this value is not optimal, with it conjectured by Ward that $c=1$ should be the optimal value \cite{ward1999}. In previous studies it has been observed that solutions exceed Ward's conjectured bound by approximately 20\%, which is a phenomenon found in other Skyrme models. 

No analytic solutions to the model defined by energy functional \eqref{eq::energy} are known, but it has been demonstrated in previous numerical work that the qualitative structure of solutions can be captured by fields generated via rational maps \cite{battye1998,battye1999,sutcliffe2007}. An energy minimisation scheme can then be implemented to find solutions. We begin by mapping  physical space $(x_1,x_2,x_3)\in\mathbb{R}^3$ to the unit three-sphere $(Z_1,Z_0)\in S^3\subset \mathbb{C}^2$ via
\begin{equation}\label{eq::zsphere}
(Z_1,Z_0)=\left( (x_1+ix_2)\frac{\sin f}{r}, \cos f + i\frac{\sin f}{r}x_3\right),
\end{equation}
where $r^2=x_1^2+x_2^2+x_3^2$ and $f(r)$ is a monotonically decreasing function satisfying $f(0)=\pi$ and $f(\infty)=0$. The Riemann sphere coordinate of the field is then given by rational map $W:S^3\subset \mathbb{C}^2 \to \mathbb{CP}^1$ given by
\begin{equation}\label{eq::rational_map} 
 W(Z_1,Z_0)=\frac{\phi_1 + i \phi_2}{1+\phi_3}=\frac{p(Z_1,Z_0)}{q(Z_1,Z_0)},
\end{equation}
for some polynomial functions $p$ and $q$ which have no common factors and have no roots in common for $(Z_1,Z_0)$ on the three-sphere. Since we have fixed the field at infinity it is natural to interpret the location of the soliton as the preimage of $\boldsymbol{\phi}=(0,0,-1)$, the antipodal point to $\boldsymbol{\phi}_\infty$, giving the centreline of the string-like solitons. It can be seen that this relates to $q(Z_1,Z_0)=0$ in the rational map, while the numerator $p(Z_1,Z_0)$ describes the behaviour of the field about this centreline and so is important for defining the Hopf charge of the ansatz. We also require $p(Z_1,Z_0)$ to be such that $W\to0$ as $r\to\infty$, to satisfy the boundary conditions of the field.

The topological charge generated by a given rational map of the form \eqref{eq::rational_map} can be calculated by considering a natural extension of $W$, $\widetilde{W}:\mathbb{B}^4\subset \mathbb{C}^2\to\mathbb{C}^2$ given by $(Z_1,Z_0)\mapsto (p(Z_1,Z_0),q(Z_1,Z_0))$. Here $(Z_1,Z_0)\in\mathbb{B}^4$ are given by relaxing the constraint of $(Z_1,Z_0)$ lying on $S^3$, to $|Z_1|^2 + |Z_0|^2\le1$. Consider the preimages of a regular point close to the roots of $\widetilde{W}$. Consider a three-sphere about each of these preimages, such that the roots of $\widetilde{W}$ are contained within the three-spheres. $\widetilde{W}$ imposes, up to some normalisation, a unit vector field in $S^3$ on the four-ball excluding the roots of $\widetilde{W}$. Since $p$ and $q$ are holomorphic, the map from each three-sphere to the vector field has unit topological charge. Taking a connected sum of these spheres maintains the charge and one can then deform this surface to the boundary of the four-ball. Thus the topological charge of the mapping from the $S^3$ forming the boundary of the four-ball to the vector field taking values in $S^3$ has topological charge given by the number of preimages under $\widetilde{W}$ of a regular point. The standard Hopf mapping is then used and we see that any normalisation factor in the definition of the vector field cancels, so we regain $W$. Since under the standard Hopf mapping, the Hopf charge of the map is the topological charge of the map between three-spheres, it follows that the Hopf degree of $W$ is the number of preimages inside the four-ball of a regular point under $\widetilde{W}$.

We now briefly recap types of rational map that have been used previously and introduce our notation for referring to these. Axially symmetric solutions have long been known to exist in the model, and fields with the correct structure are generated by the rational map, 
\begin{equation}\label{eq::axial}
W=\frac{Z_1^m}{Z_0^n},
\end{equation}
for $m$ and $n$ positive integers. This field describes an axial solution which winds $n$ times longitudinally and $m$ times in the meridional direction. This results in a field configuration with $Q=mn$. We shall denote such field configurations by $\mathcal{A}_{m,n}$.

A knotted field configuration of the form of an $(a,b)$-torus knot is given by 
\begin{equation}
W=\frac{Z_1^\alpha Z_0^\beta}{Z_1^a+Z_0^b},
\label{eq::torus}
\end{equation}
for $a$ and $b$ coprime, $\alpha$ a positive integer and $\beta$ a non-negative integer. We can see that this must indeed form a torus knot since it is known that the intersection of the plane curve $Z_1^a+Z_0^b=0$ and the unit three-sphere results in an $(a,b)$-torus knot \cite{brieskorn1986}. Solving $(Z_1^\alpha Z_0^\beta, Z_1^a+Z_0^b)=(\epsilon,0)$ for $Z_1$ we see that solutions obey $Z_1^{\alpha b +\beta a}=(-1)^{\beta}\epsilon^b$. All the roots lie within the unit four-ball for $\epsilon$ sufficiently small and so the topological charge of the rational map is $Q=\alpha b +\beta a$. An $(a,b)$-torus knot is equivalent to an $(b,a)$-torus knot, so we shall denote such fields by $\mathcal{K}_{a,b}$ with the convention that $a>b$. 

\begin{figure}
\centering
\begin{subfigure}[b]{0.24\textwidth}
	\centering
	\includegraphics[width=\textwidth, clip=true, trim= 20 45 50 30]{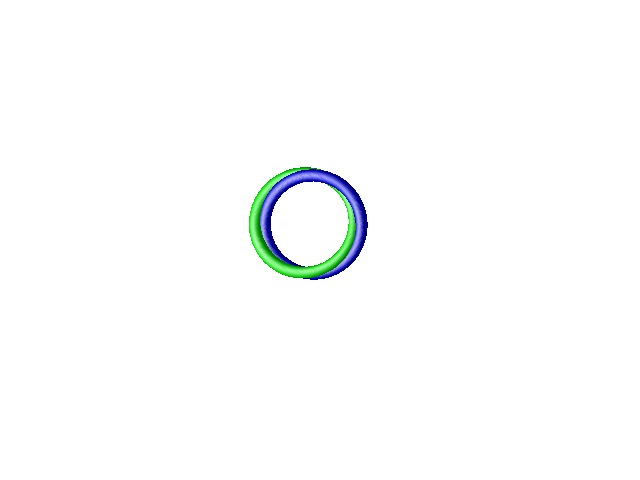}
	\caption{$1\mathcal{A}_{1,1}$}
		\label{fig::1a11}
\end{subfigure}
\begin{subfigure}[b]{0.24\textwidth}
	\centering
	\includegraphics[width=\textwidth, clip=true, trim= 35 35 35 35]{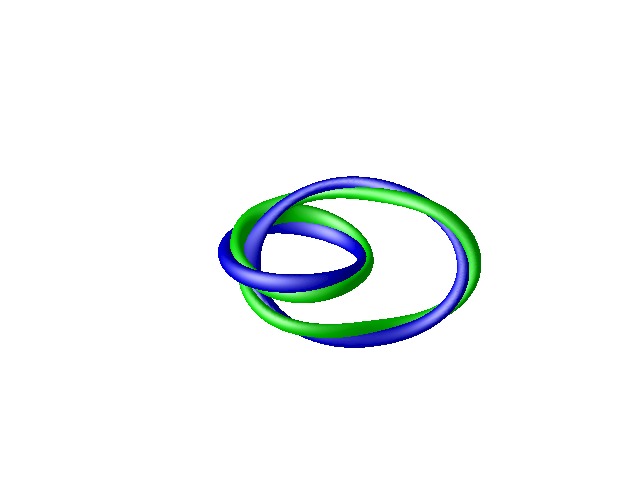}
	\caption{$5\mathcal{L}_{2,1}$}
		\label{fig::5l21}
\end{subfigure}
\begin{subfigure}[b]{0.24\textwidth}
	\centering
	\includegraphics[width=\textwidth, clip=true, trim= 35 35 35 35]{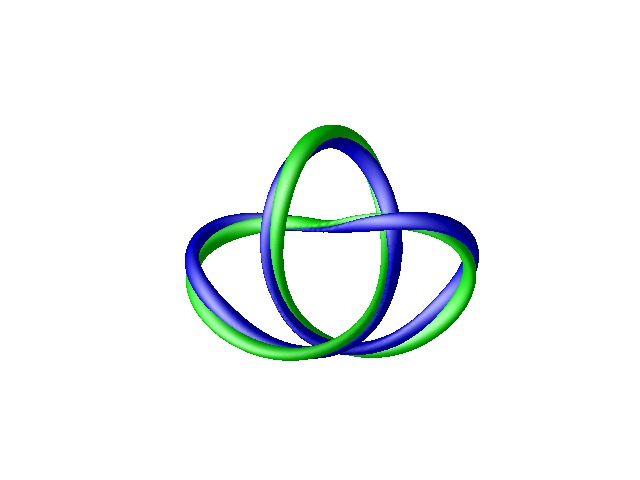}
	\caption{$7\mathcal{K}_{3,2}$}
		\label{fig::7k32}
\end{subfigure}
\begin{subfigure}[b]{0.24\textwidth}
	\centering
	\includegraphics[width=\textwidth, clip=true, trim= 35 35 35 35]{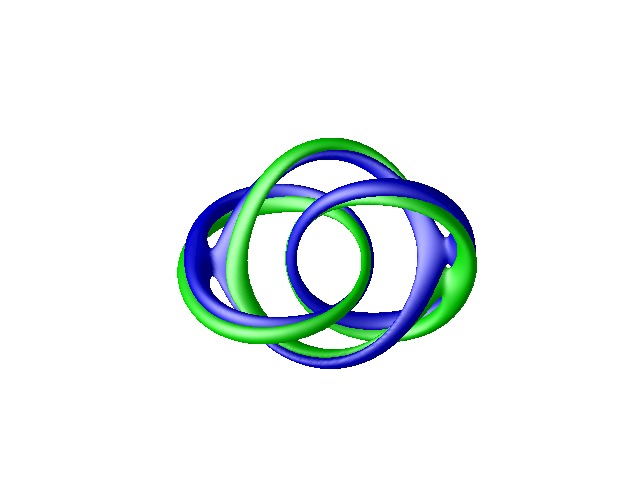}
	\caption{$9\mathcal{L}_{1,1,1}^{2,2,2}$}
		\label{fig::9l111}
\end{subfigure}
\caption{A selection of well known lower charge solutions demonstrating know forms of solutions. Position curves (blue) and linking curve (green). }
\label{fig::lowcharge}
\end{figure}

A rational map with denominator that is reducible into distinct factors generates a field with multiple components, each relating to a factor. This allows us to construct a field that takes the form of a link. For example a link of $\mathcal{A}_{m,1}$ and $\mathcal{A}_{n,1}$ is given by rational map
\begin{equation}\label{eq::linkexample}
W= \frac{Z_1^{m+1} + Z_1^{n+1}+Z_0(Z_1^m-Z_1^n)}{2(Z_1^2-Z_0^2)} =\frac{Z_1^m}{2(Z_1-Z_0)} + \frac{Z_1^n}{2(Z_1 + Z_0)},
\end{equation}
where we see that each of the constituent parts are of the form of \eqref{eq::axial} after a suitable transformation of the denominator to form the correct linking behaviour. We shall denote this link structure for torus knots as $\mathcal{L}^{\ell_1, \ell_2,\cdots}_{Q_1(a_1,b_1),Q_2(a_2,b_2),\cdots}$. Since we take preimages of points which are close, one can naturally distinguish a pair of curves per component of the link. The subscript gives the linking number $Q_i$ of this pair of curves for each component with form $\mathcal{K}_{a_i,b_i}$. The corresponding superscript then gives the additional charge gained by linking of the location curve with the second preimage of other components. Thus the Hopf charge of the link is given by summing the $Q_i$ and $\ell_i$. For example, \eqref{eq::linkexample} gives $\mathcal{L}_{m(1,1),n(1,1)}^{1,1}$, since each component is of charge $m$ and $n$ respectively, and since each position curve is linked once with the other. However, in the case where components are trivial $(1,1)$ torus knots, so axial of the form $\mathcal{A}_{m,1}$, we shall shorten our notation to denote these as $\mathcal{L}_{m,n}^{1,1}$, and so coincide with the notation of \cite{sutcliffe2007}, in which more examples of rational maps to generate links can be found. Note also that if $a$ and $b$ in \eqref{eq::torus} are not coprime the torus knot construction degenerates to form a link. A selection of low-charge hopfions of the form described by these rational maps are shown in Figure \ref{fig::lowcharge}.

\section{Cable knots and links}\label{sec::cable}
\begin{figure}
\centering
\begin{subfigure}[b]{0.4\textwidth}
	\centering	
	\includegraphics[width=0.5\textwidth]{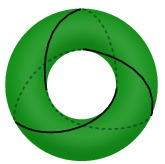}
	\caption{The $(2,3)$-torus knot (the trefoil) on the surface of a torus.}	\label{fig::cabsketcha}
\end{subfigure}
\hspace{0.1\textwidth}
\begin{subfigure}[b]{0.4\textwidth}
	\centering
	\includegraphics[width=0.5\textwidth]{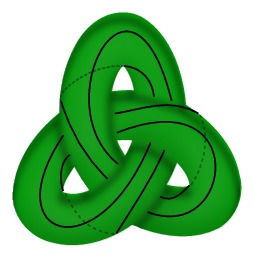}
	\caption{The $\mathcal{C}^{2,3}_{3,2}$ cable knot, lying on a neighbourhood of a trefoil.}	\label{fig::cabsketchb}
\end{subfigure}
\caption{Sketches of construction of cable knots}
\label{fig::cabsketch}
\end{figure}

We now turn our attention to finding rational maps which describe fields with the form of cable knots. These are a generalisation of torus knots, and so are a good candidate to appear in the Skyrme-Faddeev model. Given an $(m_1,n_1)$-torus knot, we can define a tubular neighbourhood about this knot. One can then embed the torus on which an $(m_2,n_2)$-torus knot lies onto this neighbourhood. The resulting knot formed by this embedding is called the $(n_2, m_2)$ cable on the $(m_1,n_1)$-torus knot. We shall denote such a field $\mathcal{C}^{n_2,m_2}_{m_1,n_1}$. Note that the ordering of parameters is now important, unlike with torus knots in general $\mathcal{C}^{n_2,m_2}_{m_1,n_1}$ is not equivalent to $\mathcal{C}^{m_2,n_2}_{m_1,n_1}$. In Figure \ref{fig::cabsketch} we sketch the construction of $\mathcal{C}^{2,3}_{3,2}$. In Figure \ref{fig::cabsketchb} we see, in green, the thickened neighbourhood of a trefoil knot. We embed the torus of a second trefoil knot, shown in Figure \ref{fig::cabsketcha}, onto this resulting in knot $\mathcal{C}^{2,3}_{3,2}$ drawn in black in \ref{fig::cabsketchb}. We have assumed above that each pair $(m_i, n_i)$ are coprime since they label torus knots, however if we relax this condition such that $(m_2,n_2)$ are no longer coprime then we find a link in the neighbourhood of the $(m_1,n_1)$-torus knot, which are called cable links.

We now wish to construct a field that is knotted in such a way. Recall that an algebraic link can be described as the intersection of  $S^3_\xi$, the three-sphere of radius $\xi>0$, with the algebraic plane curve $C=\{(x,y)\in \mathbb{C}^2: F(x,y)=0\}$, for some complex polynomial $F(x,y)$ vanishing at $(0,0)\in\mathbb{C}^2$ and under certain conditions on $\xi$. If $\xi=1$ is sufficient we merely need to set $q(Z_1,Z_0)=F(Z_1,Z_0)$. Given an $F(x,y)$ we can study the Puiseux expansion for $y$, that is writing $y$ in terms of a fractional power series in $x$ using $F(x,y)=0$, from which the type of knot this will generate can be found. However, we wish to start from such an expansion and generate the corresponding irreducible complex polynomial which it relates to. This is easily achieved following the prescription of \cite{eisenbud1985}. Briefly, to do this we rewrite our complex variables such that $x,y\in\mathbb{C}[t]$. For a specific choice of polynomials (in this study we restrict to those of the lowest degree since higher degree solutions will just give rise to the complex polynomial $F(x,y)$ for the lowest degree raised to an integer power) we then have
\begin{equation}
F(x,y)=\det (y I_n - V)
\end{equation}
where $n=\deg(x)$, $I_n$ is the $n\times n$ identity matrix and $V(x)$ is the $n\times n$ matrix defined by
\begin{equation}
yt^{i-1}= \sum_{j=1}^{n} V_{ij}t^{j-1}.
\end{equation}
To show that an expansion generates a particular knot we resort to an approach first used by K\"{a}hler \cite{kahler1929} and consider instead $C\cap D_\xi$, where $D_\xi=\{(x,y)\in\mathbb{C}^2 : |x|=\xi, |y|\le \xi\}$. It can be shown \cite{wall2004} that a sufficient condition for $C\cap D_\xi$ to be isomorphic to $C\cap S_\xi^3$ is for the vector $(x,y)$ to be nowhere orthogonal to all the tangent vectors on $C$ restricted to the four ball of radius $\xi$ with the origin removed, $\mathbb{B}_\xi^4\backslash\{(0,0)\}$. Using this alternative description of the knot it is much easier to see the form of the knot.

Consider the expansion
\begin{equation}\label{eq::expan1}
y=x^{m_1/n_1}+ \eta x^{m_2/n_1n_2},\
\end{equation}
for non-zero $\eta\in\mathbb{C}$ and positive integers $m_i$, $n_i$ which are pairwise coprime and satisfy $1<m_1/n_1 < m_2/n_1n_2$. It can be shown \cite{eisenbud1985} that this then generates the $(n_2, m_2)$-cable on $\mathcal{K}_{m_1,n_1}$. To see this we first note that this satisfies the isomorphism condition since the complex tangent converges to $y=0$ as we approach the origin. Thus in a small neighbourhood  of the origin the vector $(x,y)$ is nowhere orthogonal to all the tangent vectors, since along the curve $|y|\ll |x|$ near the origin. Considering the intersection with the solid torus, since $|x|=\xi$ is small we see that the leading behaviour in small $\xi$ will come from the first term, so $y\approx x^{m_1/n_1}$. This is parametrised by $(x,y)=(\xi t^{n_1},  \xi^{m_1/n_1}t^{m_1})$ for $t\in S^1\subset\mathbb{C}$. We see that the curve traverses $n_1$ times in the $x$ direction, and $m_1$ times in the $y$ direction, obviously tracing out $\mathcal{K}_{m_1,n_1}$, and so our knot lies in a neighbourhood of this. Considering the second term, we see that as the curve traverses longitudinally about $\mathcal{K}_{m_1,n_1}$ it rotates $m_2/n_2$ about the tube of the neighbourhood. Thus it traverses about this neighbourhood $n_2$ longitudinally and $m_2$ times in the meridional direction, and so describes an $(n_2,m_2)$ cable on $\mathcal{K}_{m_1,n_1}$ as claimed. For the cases we consider here it is sufficient for $\xi=1$, and so \eqref{eq::expan1} can be used to find fields with a cable knot structure. The first non-trivial knot from this is $\mathcal{C}^{2,7}_{3,2}$ with polynomials $x(t)=t^4$ and $y(t)=t^6+\eta t^7$ following from equation \eqref{eq::expan1}. This then leads to complex polynomial
\begin{equation}
F(x,y)=\det\begin{pmatrix} y & 0 & -x & -\eta x \\ -\eta x^2 & y & 0 & -x \\ -x^2 & -\eta x^2 & y & 0 \\ 0 & -x^2 & -\eta x^2 & y \end{pmatrix}= y^4-2x^3y^2 -4\eta^2x^5y+x^6 -\eta^4 x^7,
\end{equation}
and so the corresponding rational map is then
\begin{equation}\label{eq::(2,7)cab(2,3)}
W=\frac{Z_1^\alpha Z_0^\beta(Z_1-Z_0)^\gamma}{Z_0^4-2Z_1^3Z_0^2-4\eta^2Z_1^5Z_0+Z_1^6-\eta^4Z_1^7},
\end{equation}
for $\alpha$ a non-negative integer, $\beta$ a positive integer and $\gamma\in\{0,1\}$. The optional term, corresponding to $\gamma$ is included to enable fields to be generated which have odd topological charge. The constraint on integers $m_i$, $n_i$ is stronger than we would wish and results in the generation of more complex fields than we wish to consider initially. 

We would like however to find rational maps describing more simple cable knots. Consider the expansion
\begin{equation}\label{eq::expan2}
y=x^{3/2} + \eta x^{3/4},
\end{equation}
for non-zero $\eta\in\mathbb{C}$. This obviously does not satisfy the conditions on equation \eqref{eq::expan1}, so we need to consider the isomorphy condition again. Although in this case the tangent approaches a vector orthogonal to $(x,y)$, since the origin of the four ball is excluded, this does not forbid the isomorphy. By taking a parametrisation of the expansion one can explicitly check that it holds for $\xi>1$. In this regime, an expansion in $1/\xi$ shows the leading behaviour is still given by $\mathcal{K}_{3,2}$, with the curve lying in a neighbourhood of this knot. Applying the same logic as above we see this gives a $\mathcal{C}^{2,3}_{3,2}$ knot. We find that $\xi=1$ is again sufficient, so with polynomials $x(t)=t^4$ and $y(t)=t^6+\eta t^3$ following from equation \eqref{eq::expan2}. This then leads to complex polynomial
\begin{equation}
F(x,y)=\det\begin{pmatrix} y & 0 & -x & -\eta  \\ -\eta x & y & 0 & -x \\ -x^2 & -\eta x& y & 0 \\ 0 & -x^2 & -\eta x & y \end{pmatrix}= y^4-2x^3y^2 -4\eta^2x^3y+x^6 -\eta^4 x^3,
\end{equation}
giving the rational map which generates such a knotted field as
\begin{equation}\label{eq::(2,3)cab(2,3)}
W= \frac{Z_1^\alpha Z_0^\beta(Z_1-Z_0)^{\gamma}}{Z_0^4-2Z_1^3Z_0^2-4\eta^2Z_1^3 Z_0 + Z_1^6 -\eta^4 Z_1^3},
\end{equation}
for $\alpha$ a non-negative integer, $\beta$ a positive integer and $\gamma\in\{0,1\}$. Similar arguments also hold for constructing fields with similar cable knot structure not satisfying the condition for \eqref{eq::expan1}. When $m_2,n_2$ are not coprime, one gets cable links. These can be constructed as the sum of two components each having the form of cable knots with $m_2,n_2$ not coprime, where they differ by a suitable shift in the complex parameter $\eta$.

Finally, before presenting the results of implementing such knotted fields as initial conditions, we wish to comment on the topological charge which rational maps \eqref{eq::(2,7)cab(2,3)} and \eqref{eq::(2,3)cab(2,3)} generate. Using the approach of counting the preimages of the related mapping, one finds $Z_1$ must satisfy a polynomial equation of a particular degree. One might na\"{i}vely expect this to be the topological charge, however some of these polynomial roots scale inversely with $|\epsilon|$, and lie outside the unit ball so need to be excluded. This means that although the charge can be calculated for any given $\alpha$, $\beta$ and $\gamma$, a generic formula for the charge is complicated. The case for $\gamma=0$ can be derived analytically by considering related maps in the same homotopy class such that the roots of the polynomials remain within/outside the four-ball\footnote{The author would like to thank Martin Speight for pointing out this calculation of Hopf charge for these mappings.}. Take rational map \eqref{eq::(2,3)cab(2,3)} for example and consider the roots of
\begin{equation}
(Z_1^\alpha Z_0^\beta, Z_0^4-2Z_1^3Z_0^2 -4s\eta^2Z_1^3Z_0+Z_1^6 -s\eta^4 Z_1^3)
\end{equation}
for $s\in[0,1]$. We see that these have roots where $Z_1=Z_0=0$ or $Z_0=0$ and $Z_1^3=s\eta^4$. For $|\eta|<1$ these roots lie within the four-ball. Thus we can take $s\to0$. Now consider the mapping
\begin{equation}
(Z_1^\alpha Z_0^\beta, Z_0^4-2sZ_1^3Z_0^2 +Z_1^6)
\end{equation}
for $s\in[0,1]$. All roots now occur at $Z_1=Z_0=0$ and so we can again take $s\to0$. Thus \eqref{eq::(2,3)cab(2,3)} with $\gamma=0$ is in the same homotopy class as a particular torus knot, and so has charge $4\alpha+6\beta$. Similar arguments are satisfied by the other rational maps used with $\gamma=0$, and again it is found that $Q=4\alpha + 6\beta$.

\begin{figure}
\begin{subfigure}[b]{0.45\textwidth}
	\centering
	\includegraphics[width=\textwidth]{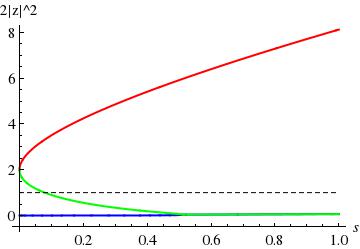}
	\caption{$\eta=0.5$}
	\label{fig::rootspoint5}
\end{subfigure}
\qquad
\begin{subfigure}[b]{0.45\textwidth}
	\centering
	\includegraphics[width=\textwidth]{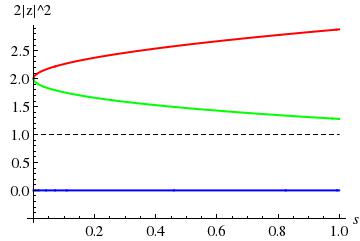}
	\caption{$\eta=0.1$}
	\label{fig::rootspoint1}
\end{subfigure}
\caption{Plots of $2|z|^2$ for the roots to $z^3-2z^2 + (1-4s\eta^2)z-s\eta^4=0$ for $s\in[0,1]$ for different values of $\eta$. Also marked in the value at which the root lies on the three-sphere.}
\label{fig::roots}
\end{figure}

For $\gamma=1$ we can consider a similar argument. Again using the example of \eqref{eq::(2,3)cab(2,3)}, consider the roots of
\begin{equation}
(Z_1^\alpha Z_0^\beta(Z_1-Z_0), Z_0^4-2Z_1^3Z_0^2 -4s\eta^2Z_1^3Z_0+Z_1^6 -s\eta^4 Z_1^3)
\end{equation}
for $s\in[0,1]$. We see that these have roots where $Z_1=Z_0=0$, $Z_0=0$ and $Z_1^3=s\eta^4$ or $Z_1=Z_0$ and $Z_1^3-2Z_1^2 + (1-4s\eta^2)Z_1-s\eta^4=0$. For choices of $\eta$ small enough, such that these roots remain inside/outside the three-sphere, we can then take $s\to0$ and remain in the same homotopy class. Consider 
\begin{equation}
(Z_1^\alpha Z_0^\beta(Z_1-sZ_0), Z_0^4-2Z_1^3Z_0^2 +Z_1^6)
\end{equation}
for $s\in[0,1]$. Roots now occur for $Z_1=Z_0=0$ or $Z_1=sZ_0$ and $1-2s^3Z_0 + s^6Z_0^2=0$. These roots remain inside/outside the four-ball as $s\to0$. Finally consider
\begin{equation}
(Z_1^{\alpha+1} Z_0^\beta, Z_0^4-2sZ_1^3Z_0^2 +Z_1^6)
\end{equation}
for $s\in[0,1]$. All roots lie at $Z_1=Z_0=0$ and so can take $s\to0$. Thus we again find that if our choice of $\eta$ is sufficient, then \eqref{eq::(2,3)cab(2,3)} with $\gamma=1$ is in the same homotopy class as a particular torus knot and so has charge $4(\alpha+1)+6\beta$. It remains to determine whether the choice of $\eta$ is sufficiently small. In this study we use $\eta=1/2$. We find that our choice does not satisfy this condition. As can be seen in Figure \ref{fig::rootspoint5} in such a limit one of the roots cross the boundary of the four-ball, and so the above argument does not hold. For a different choice of $\eta$ which is small enough, as in Figure \ref{fig::rootspoint1} where $\eta=0.1$, this argument would hold. Instead we resort to numerically calculating the number of preimages within the four-ball and find that for $\eta=1/2$ this map has charge $4\alpha+6\beta+5$. A similar argument also holds for the other mappings considered here.

\section{Numerical Results}\label{sec::results}
The numerical simulations have been carried out on a cubic grid with $(301)^3$ grid points and with lattice spacing $dx=0.06$. Points on the edge of the grid are fixed to be $\boldsymbol{\phi}_\infty$. The energy of a given initial field as described in the previous sections is then minimised by the approach of \cite{sutcliffe2007}. The field is evolved according to field equations derived from \eqref{eq::energy}, where only the second order kinetic terms from the sigma model term are considered. Whenever the potential energy of the system increases the kinetic energy is removed. This approach results in an algorithm which converges faster than standard gradient flow algorithms whilst avoiding the numerically costly matrix inversion which the complete dynamical field equations would entail. Limitations of the numerics also required us to scale $W$ in some of the initial conditions so the field changes more smoothly.

We present our solutions by plotting an isosurface where $\phi_3=-0.95$ which is the boundary of a tubular neighbourhood of the position of the soliton. We also plot a neighbourhood of a point close to the position curve to encapsulate the twisting of the field, and so enable us to see the Hopf charge as the linking number of these two curves. We choose this second point to be $\boldsymbol{\phi}=(\sqrt{2\mu-\mu^2},0, \mu-1)$ for $\mu=0.1$ and take an isosurface about this value. In some cases, the isosurface plotted interacts and makes it difficult to determine the form of the solution. In these cases to resolve the curve it was necessary to consider preimages of $\phi_3=-1$ as the isoline where $\phi_1=\phi_2=0$ where we restrict to field values on the southern hemisphere of the target $S^2$ to aid interpolation.

\begin{figure}
\begin{subfigure}[b]{0.19\textwidth}
	\includegraphics[width=\textwidth, clip=true, trim= 35 0 35 0]{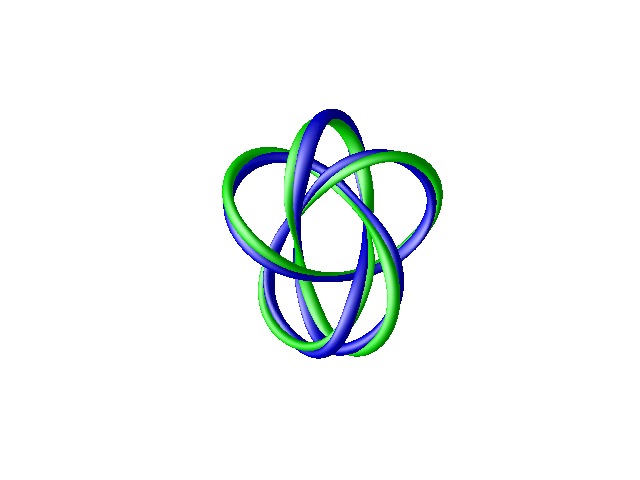}
	\caption{$16\mathcal{K}_{5,3}$}
	\label{fig::16k53}
\end{subfigure}
\begin{subfigure}[b]{0.19\textwidth}
	\includegraphics[width=\textwidth, clip=true, trim= 35 0 35 0]{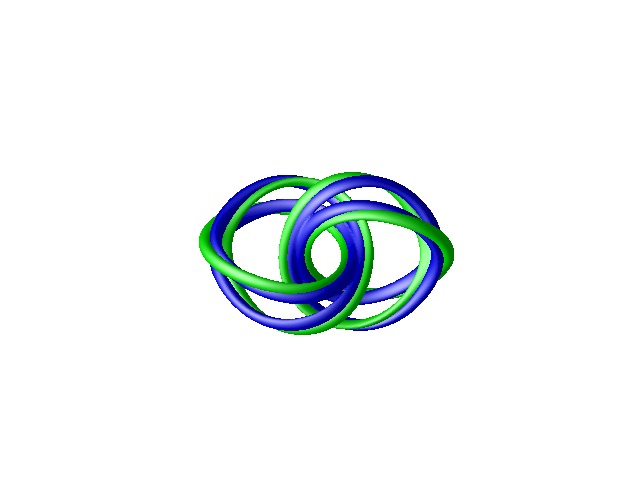}
	\caption{$16\mathcal{L}_{4,4}^{4,4}$}
	\label{fig::16l44}
\end{subfigure}
\begin{subfigure}[b]{0.19\textwidth}
	\includegraphics[width=\textwidth, clip=true, trim= 35 0 35 0]{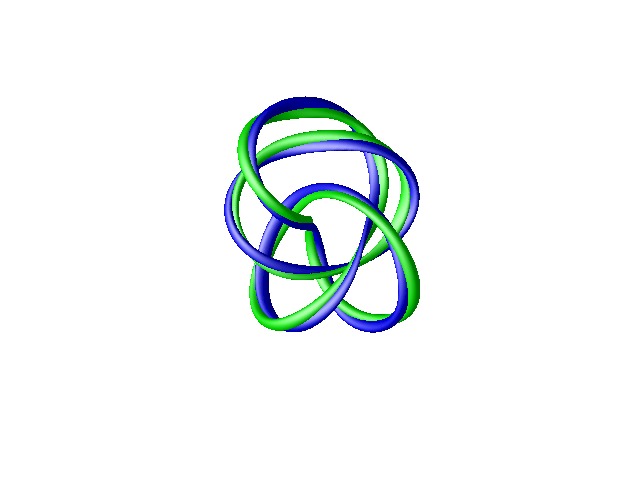}
	\caption{$16\mathcal{L}_{8(3,2),2}^{3,3}$}
	\label{fig::16l82}
\end{subfigure}
\begin{subfigure}[b]{0.19\textwidth}
	\includegraphics[width=\textwidth, clip=true, trim= 35 0 35 0]{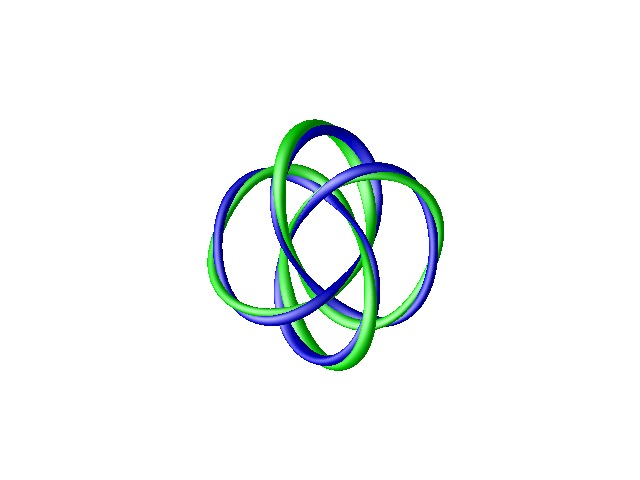}
	\caption{$16\mathcal{K}_{4,3}$}
	\label{fig::16k43}
\end{subfigure}
\begin{subfigure}[b]{0.19\textwidth}
	\includegraphics[width=\textwidth, clip=true, trim= 35 0 35 0]{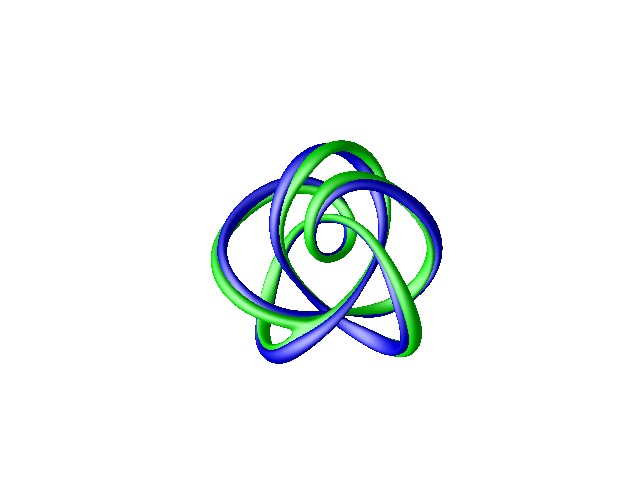}
	\caption{$16\mathcal{L}_{5(3,2),3}^{4,4}$}
	\label{fig::16l53}
\end{subfigure}
\begin{subfigure}[b]{0.19\textwidth}
	\includegraphics[width=\textwidth, clip=true, trim= 35 0 35 0]{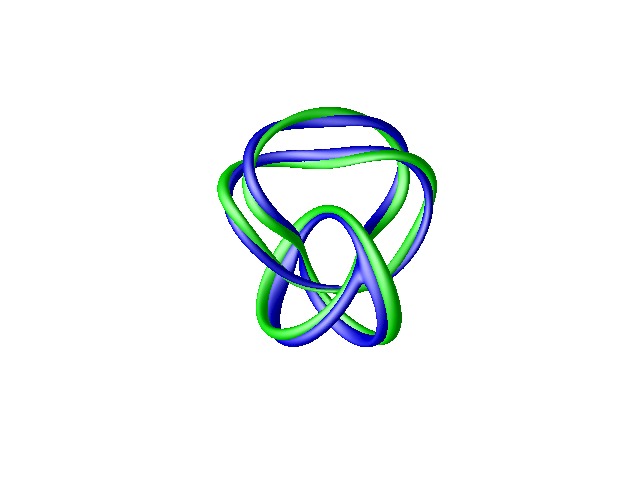}
	\caption{$16\mathcal{L}_{7(3,2),3}^{3,3}$}
	\label{fig::16l73}
\end{subfigure}
\begin{subfigure}[b]{0.19\textwidth}
	\includegraphics[width=\textwidth, clip=true, trim= 35 0 35 0]{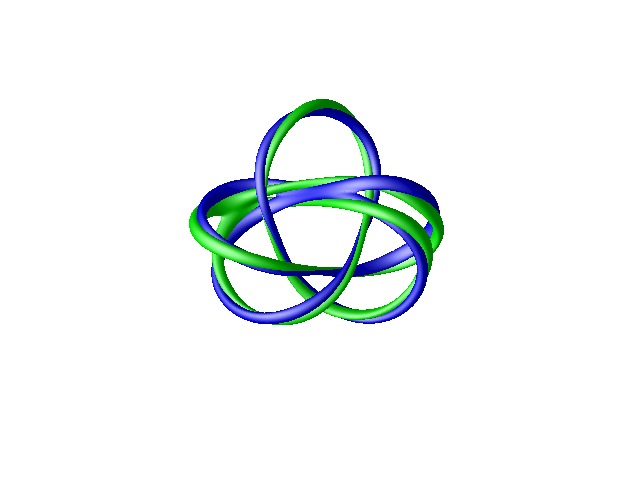}
	\caption{$17\mathcal{L}_{8(3,2),3}^{3,3}$}
	\label{fig::17l83}
\end{subfigure}
\begin{subfigure}[b]{0.19\textwidth}
	\includegraphics[width=\textwidth, clip=true, trim= 35 0 35 0]{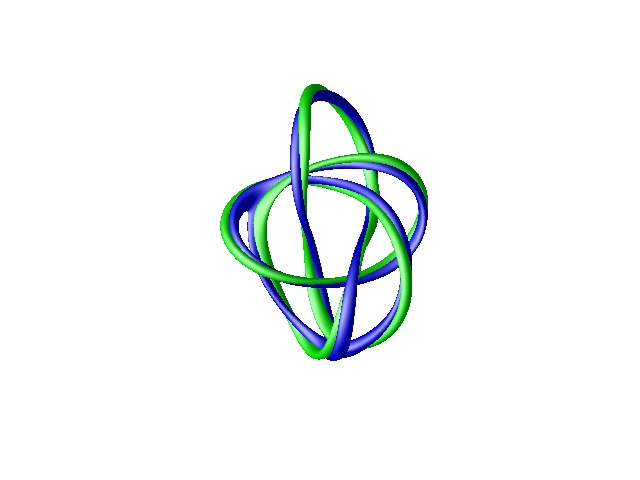}
	\caption{$17\mathcal{L}_{9(3,2),2}^{3,3}$}
	\label{fig::17l92}
\end{subfigure}
\begin{subfigure}[b]{0.19\textwidth}
	\includegraphics[width=\textwidth, clip=true, trim= 35 0 35 0]{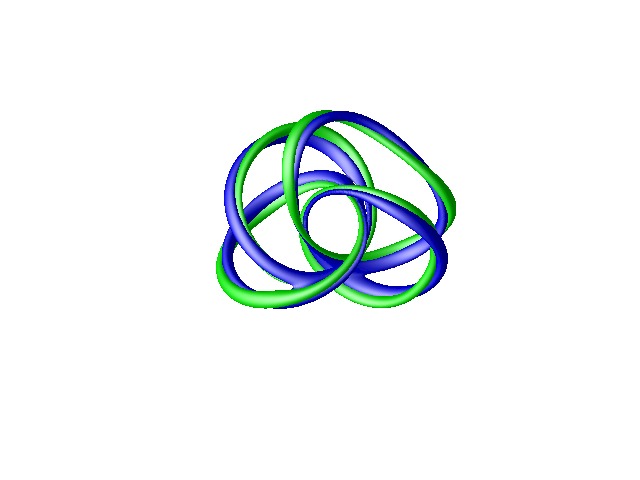}
	\caption{$17\mathcal{L}_{2,1,1,1}^{3,3,3,3}$}
	\label{fig::17l2111}
\end{subfigure}
\begin{subfigure}[b]{0.19\textwidth}
	\includegraphics[width=\textwidth, clip=true, trim= 35 0 35 0]{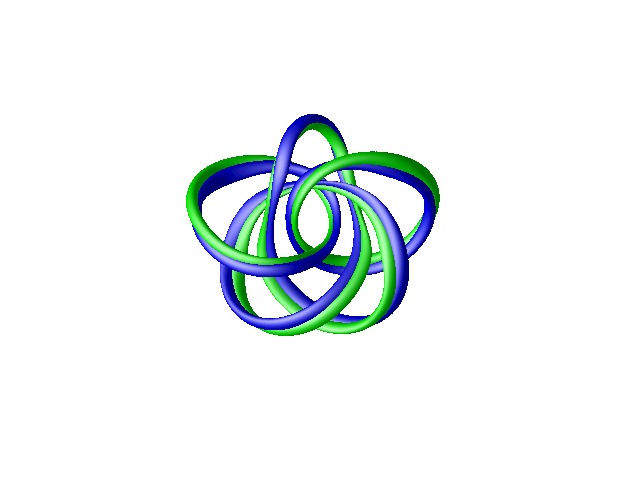}
	\caption{$17\mathcal{L}_{5(3,2),1,1}^{4,3,3}$}
	\label{fig::17l511}
\end{subfigure}
\begin{subfigure}[b]{0.19\textwidth}
	\includegraphics[width=\textwidth, clip=true, trim= 35 0 35 0]{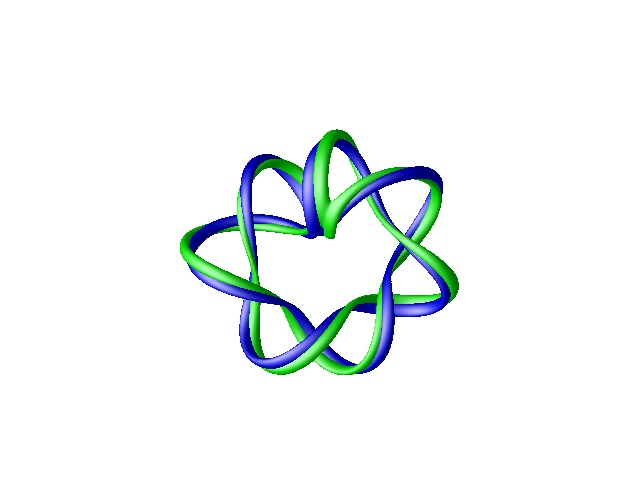}
	\caption{$17\mathcal{K}_{7,2}$}
	\label{fig::17k72}
\end{subfigure}
\begin{subfigure}[b]{0.19\textwidth}
	\includegraphics[width=\textwidth, clip=true, trim= 35 0 35 0]{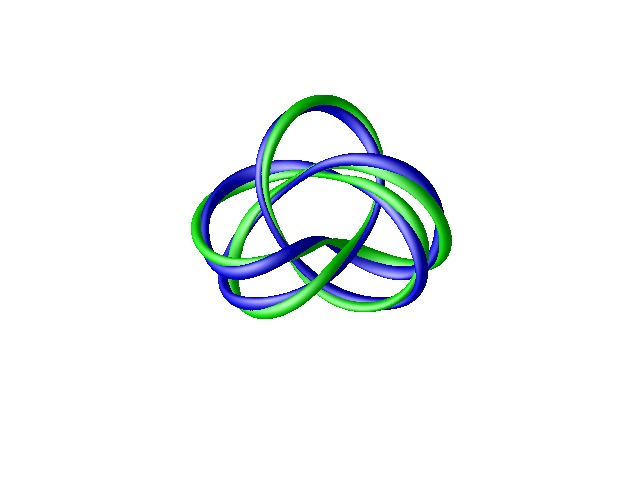}
	\caption{$18\mathcal{K}_{5,3}$}
	\label{fig::18k53}
\end{subfigure}
\begin{subfigure}[b]{0.19\textwidth}
	\includegraphics[width=\textwidth, clip=true, trim= 35 0 35 0]{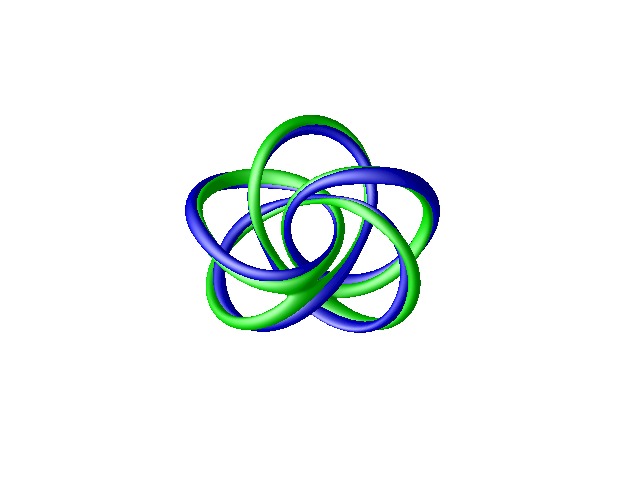}
	\caption{$18\mathcal{L}_{6(3,2), 1,1}^{ 4,3,3}$}
	\label{fig::18l611}
\end{subfigure}
\begin{subfigure}[b]{0.19\textwidth}
	\includegraphics[width=\textwidth, clip=true, trim= 35 0 35 0]{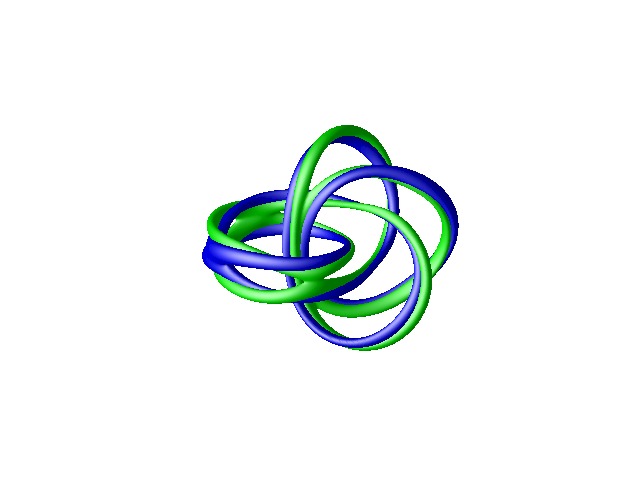}
	\caption{$18\mathcal{L}_{11(5,2), 1}^{3,3}$}
	\label{fig::18l111}
\end{subfigure}
\begin{subfigure}[b]{0.19\textwidth}
	\includegraphics[width=\textwidth, clip=true, trim= 35 0 35 0]{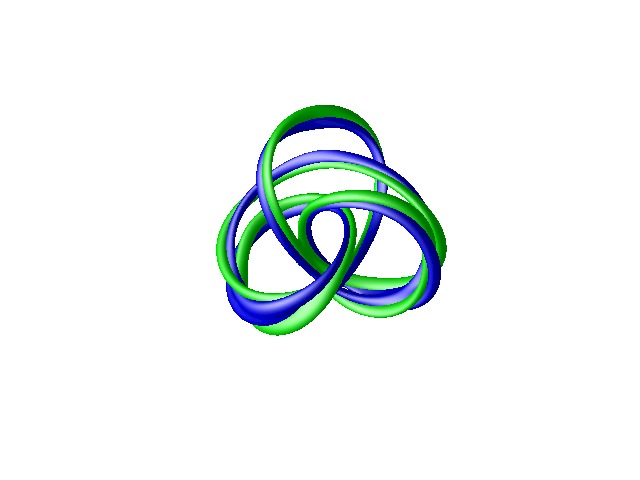}
	\caption{$\boldsymbol{18\mathcal{H}_{4.851}}$}
	\label{fig::18h}
\end{subfigure}
\begin{subfigure}[b]{0.19\textwidth}
	\includegraphics[width=\textwidth, clip=true, trim= 35 0 35 0]{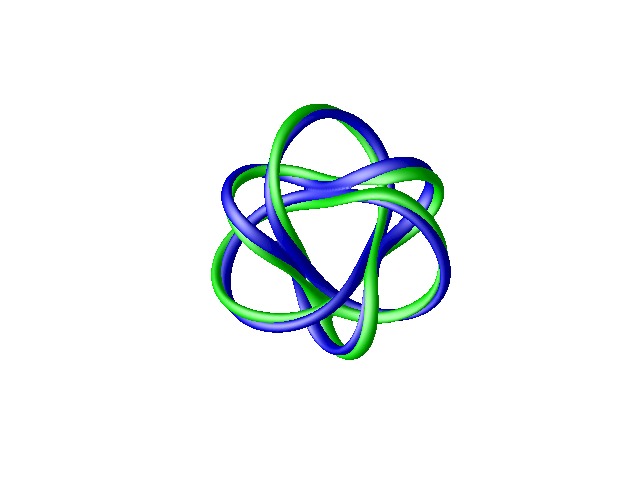}
	\caption{$18\mathcal{L}_{8(3,2), 4}^{3,3}$}
	\label{fig::18l84}
\end{subfigure}
\begin{subfigure}[b]{0.19\textwidth}
	\includegraphics[width=\textwidth, clip=true, trim= 35 0 35 0]{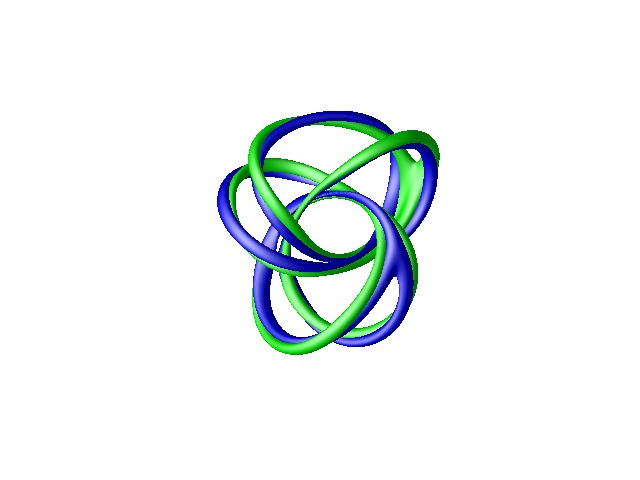}
	\caption{$19\mathcal{K}_{5,4}$}
	\label{fig::19k54}
\end{subfigure}
\begin{subfigure}[b]{0.19\textwidth}
	\includegraphics[width=\textwidth, clip=true, trim= 35 0 35 0]{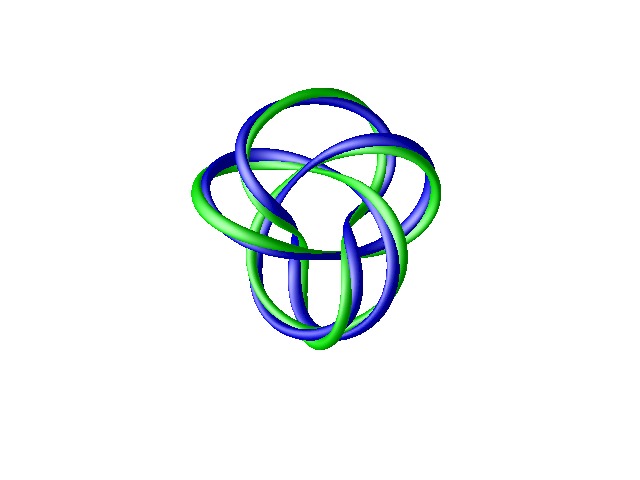}
	\caption{$19\mathcal{K}_{5,3}$}
	\label{fig::19k53}
\end{subfigure}
\begin{subfigure}[b]{0.19\textwidth}
	\includegraphics[width=\textwidth, clip=true, trim= 35 0 35 0]{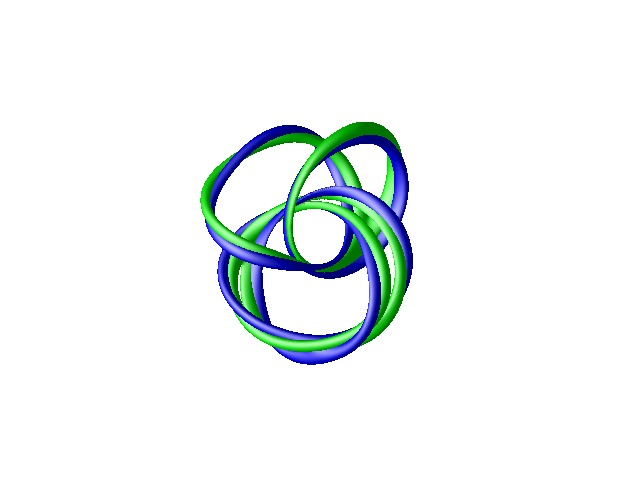}
	\caption{$19\mathcal{L}_{6(3,2),2,1}^{4,3,3}$}
	\label{fig::19l621}
\end{subfigure}
\begin{subfigure}[b]{0.19\textwidth}
	\includegraphics[width=\textwidth, clip=true, trim= 35 0 35 0]{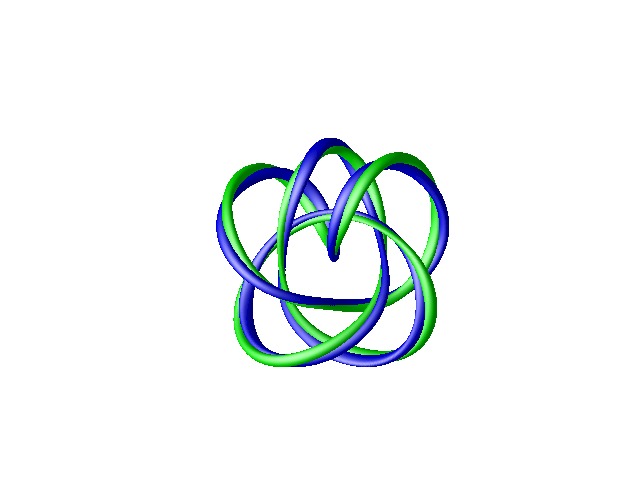}
	\caption{$19\mathcal{L}_{7(3,2),4}^{4,4}$}
	\label{fig::19l74}
\end{subfigure}
\begin{subfigure}[b]{0.19\textwidth}
	\includegraphics[width=\textwidth, clip=true, trim= 35 0 35 0]{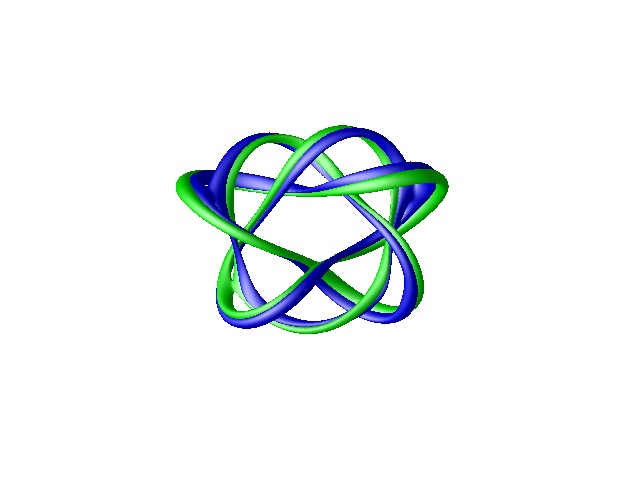}
	\caption{$19\mathcal{K}_{7,3}$}
	\label{fig::19k73}
\end{subfigure}
\begin{subfigure}[b]{0.19\textwidth}
	\includegraphics[width=\textwidth, clip=true, trim= 35 0 35 0]{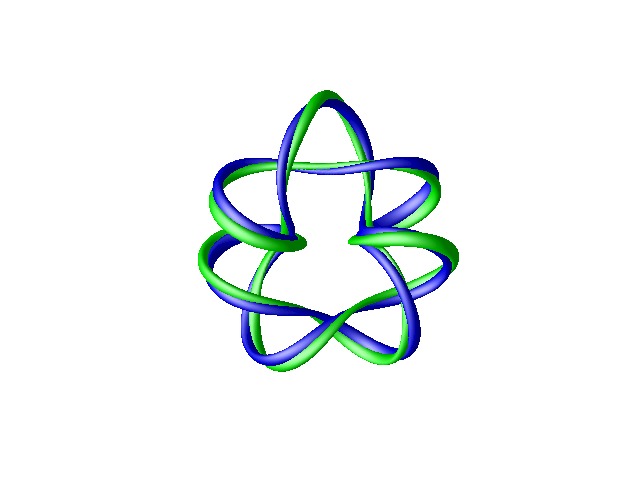}
	\caption{$19\mathcal{K}_{7,2}$}
	\label{fig::19k72}
\end{subfigure}
\caption{The position curves (blue) and linking curve (green) for a range of solutions with topological charge $16\le Q \le 19$. Solutions not formed of torus knots are marked in bold.}
\label{fig::mainresult1} 
\end{figure}

\begin{figure}
\begin{subfigure}[b]{0.19\textwidth}
	\includegraphics[width=\textwidth, clip=true, trim= 35 0 35 0]{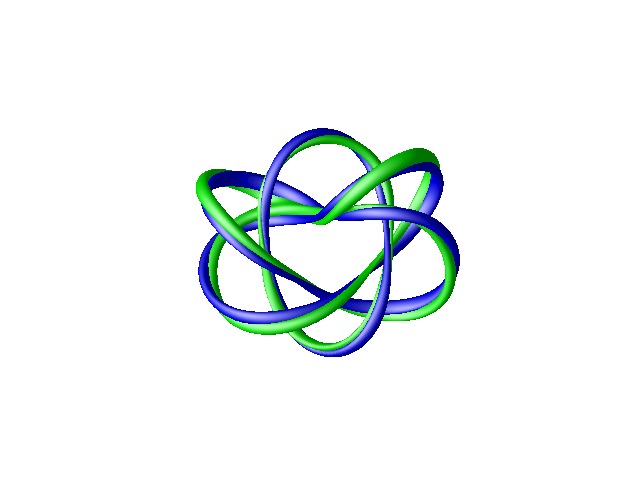}
	\caption{$20\mathcal{L}^{4,4,4}_{3,3,2}$}
	\label{fig::20l332}
\end{subfigure}
\begin{subfigure}[b]{0.19\textwidth}
	\includegraphics[width=\textwidth, clip=true, trim= 35 0 35 0]{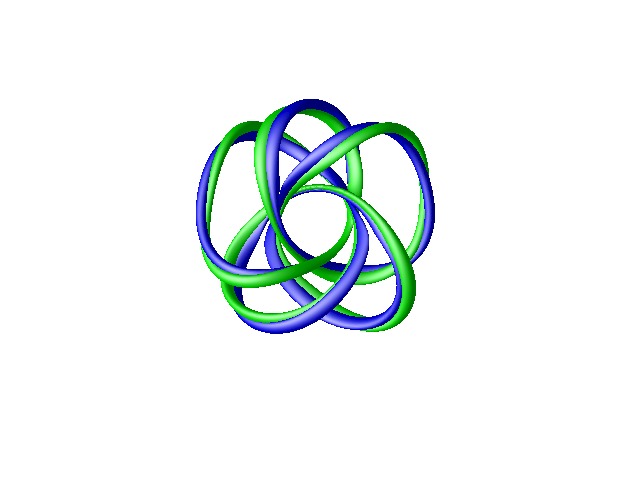}
	\caption{$20\mathcal{K}_{5,4}$}
	\label{fig::20k54}
\end{subfigure}
\begin{subfigure}[b]{0.19\textwidth}
	\includegraphics[width=\textwidth, clip=true, trim= 35 0 35 0]{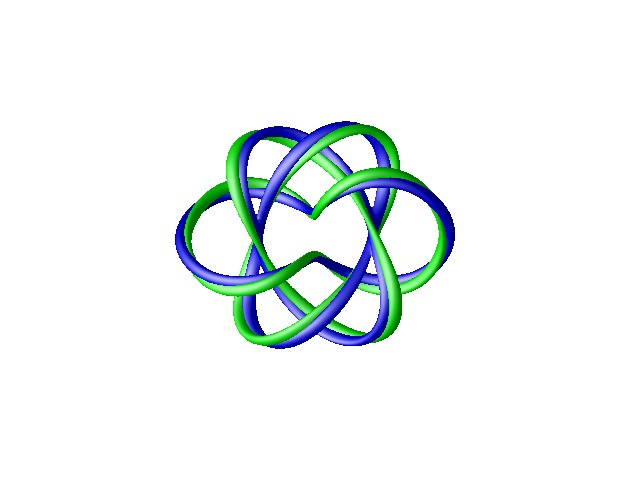}
	\caption{$20\mathcal{L}^{4,4,4}_{4,2,2}$}
	\label{fig::20l422}
\end{subfigure}
\begin{subfigure}[b]{0.19\textwidth}
	\includegraphics[width=\textwidth, clip=true, trim= 35 0 35 0]{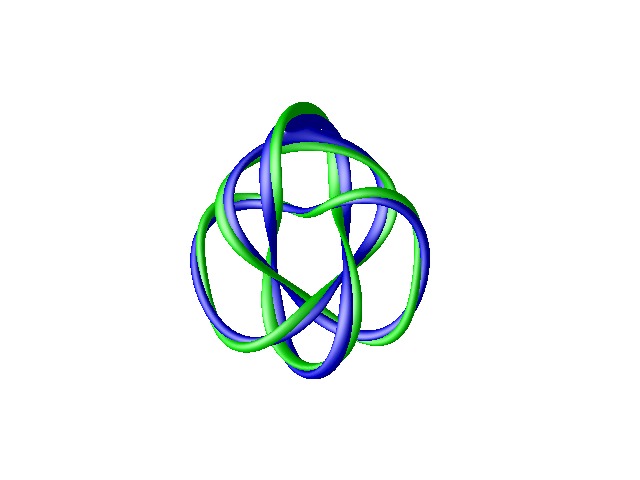}
	\caption{$20\mathcal{L}^{4,4}_{9(3,2),3}$}
	\label{fig::20l93}
\end{subfigure}
\begin{subfigure}[b]{0.19\textwidth}
	\includegraphics[width=\textwidth, clip=true, trim= 35 0 35 0]{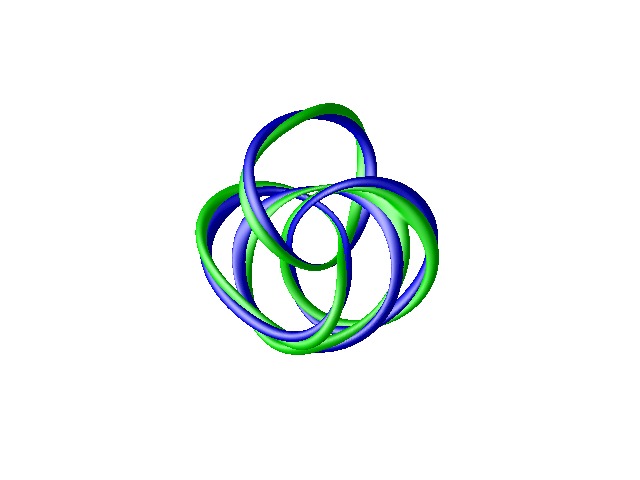}
	\caption{$20\mathcal{L}^{3,3,3,3}_{2,2,2,2}$}
	\label{fig::20l2222}
\end{subfigure}
\begin{subfigure}[b]{0.19\textwidth}
	\includegraphics[width=\textwidth, clip=true, trim= 35 0 35 0]{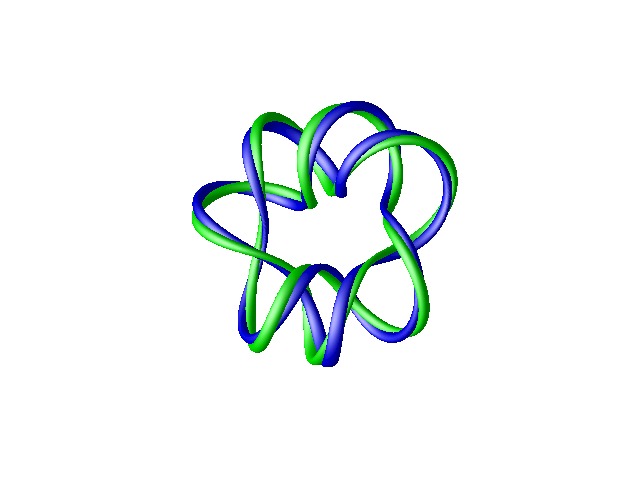}
	\caption{$20\mathcal{K}_{7,2}$}
	\label{fig::20k72}
\end{subfigure}
\begin{subfigure}[b]{0.19\textwidth}
	\includegraphics[width=\textwidth, clip=true, trim= 35 0 35 0]{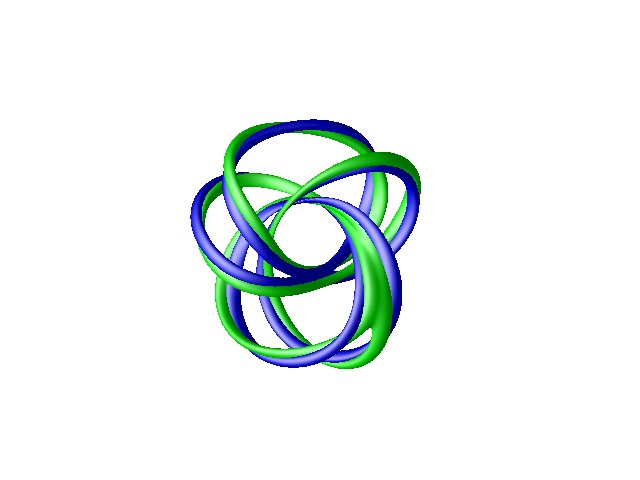}
	\caption{$21\mathcal{L}_{13(4,3),2}^{3,3}$}
	\label{fig::21l132}
\end{subfigure}
\begin{subfigure}[b]{0.19\textwidth}
	\includegraphics[width=\textwidth, clip=true, trim= 35 0 35 0]{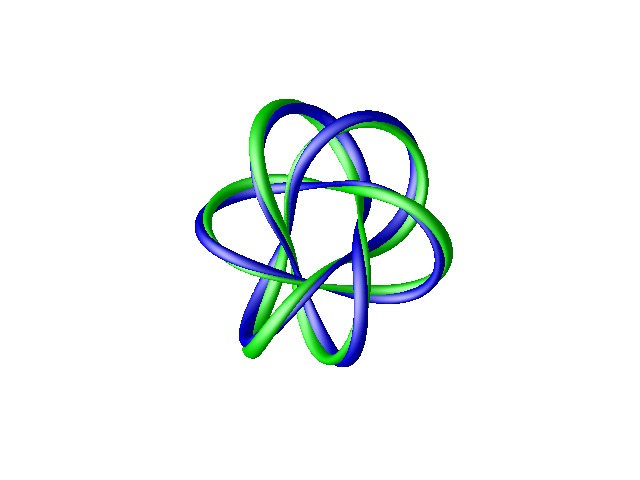}
	\caption{$21\mathcal{L}_{3,3,3}^{4,4,4}$}
	\label{fig::21l333}
\end{subfigure}
\begin{subfigure}[b]{0.19\textwidth}
	\includegraphics[width=\textwidth, clip=true, trim= 35 0 35 0]{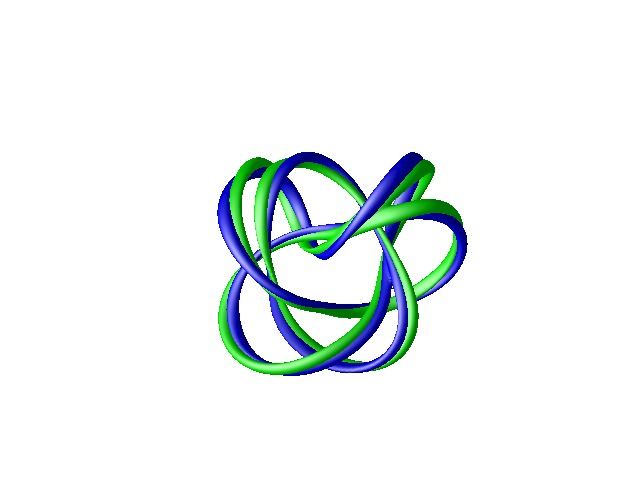}
	\caption{$21\mathcal{L}_{11(5,2),2}^{4,4}$}
	\label{fig::21l112}
\end{subfigure}
\begin{subfigure}[b]{0.19\textwidth}
	\includegraphics[width=\textwidth, clip=true, trim= 35 0 35 0]{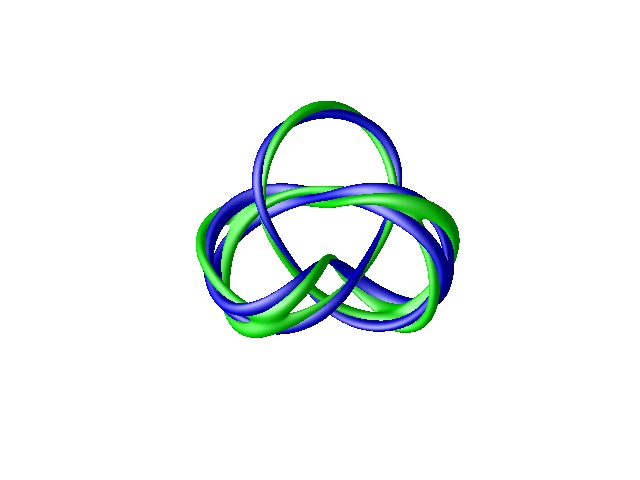}
	\caption{$21\mathcal{K}_{5,3}$}
	\label{fig::21k53}
\end{subfigure}
\begin{subfigure}[b]{0.19\textwidth}
	\includegraphics[width=\textwidth, clip=true, trim= 35 0 35 0]{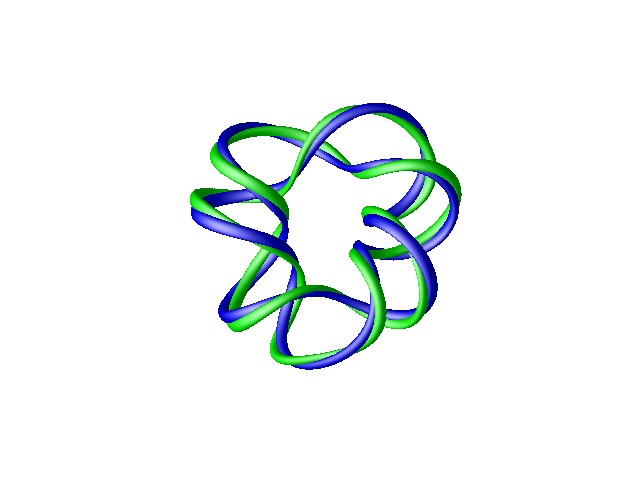}
	\caption{$21\mathcal{K}_{7,2}$}
	\label{fig::21k72}
\end{subfigure}
\begin{subfigure}[b]{0.19\textwidth}
	\includegraphics[width=\textwidth, clip=true, trim= 35 0 35 0]{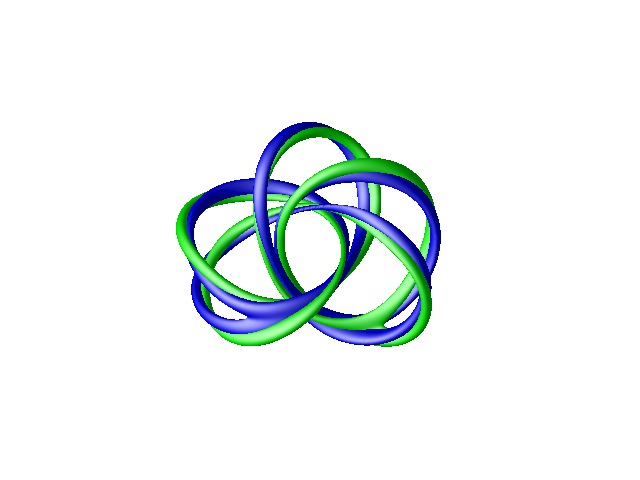}
	\caption{$22\mathcal{L}_{8(3,2),2,2}^{4,3,3}$}
	\label{fig::22l822}
\end{subfigure}
\begin{subfigure}[b]{0.19\textwidth}
	\includegraphics[width=\textwidth, clip=true, trim= 35 0 35 0]{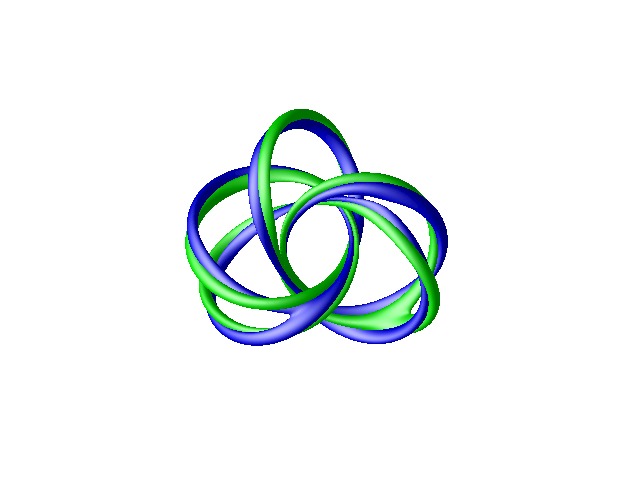}
	\caption{$22\mathcal{K}_{5,4}$}
	\label{fig::22k54}
\end{subfigure}
\begin{subfigure}[b]{0.19\textwidth}
	\includegraphics[width=\textwidth, clip=true, trim= 35 0 35 0]{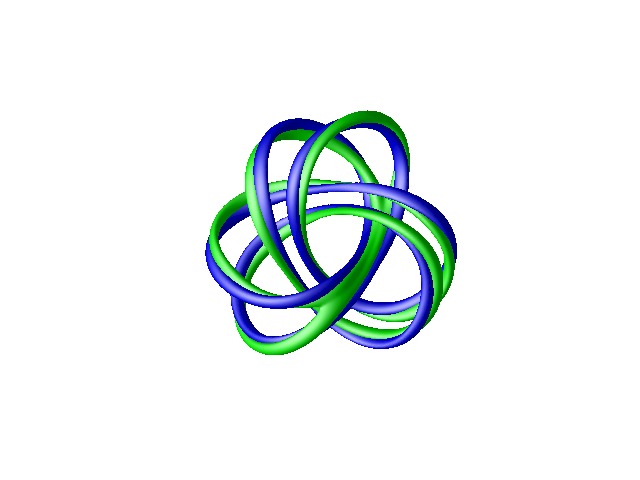}
	\caption{$\boldsymbol{22\mathcal{C}^{2,3}_{3,2}}$}
	\label{fig::22c32_32}
\end{subfigure}
\begin{subfigure}[b]{0.19\textwidth}
	\includegraphics[width=\textwidth, clip=true, trim= 35 0 35 0]{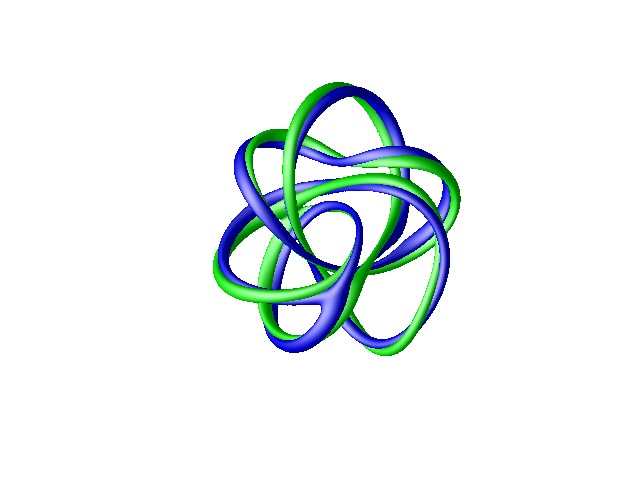}
	\caption{$22\mathcal{L}_{7(3,2),2,1}^{5,4,3}$}
	\label{fig::22l721}
\end{subfigure}
\begin{subfigure}[b]{0.19\textwidth}
	\includegraphics[width=\textwidth, clip=true, trim= 35 0 35 0]{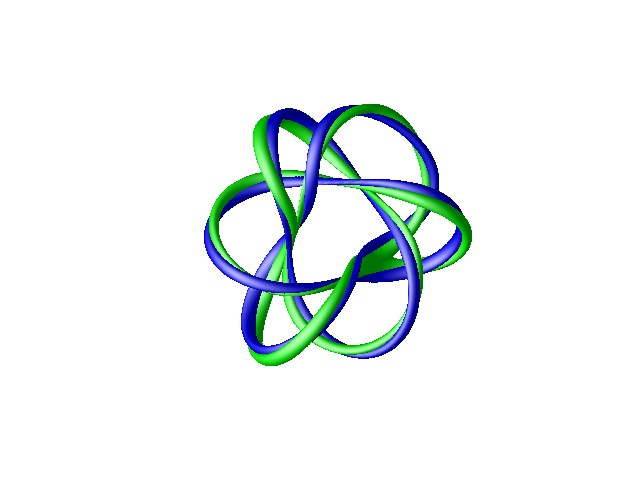}
	\caption{$22\mathcal{L}^{4,4,4}_{4,3,2}$}
	\label{fig::22l432}
\end{subfigure}
\begin{subfigure}[b]{0.19\textwidth}
	\includegraphics[width=\textwidth, clip=true, trim= 35 0 35 0]{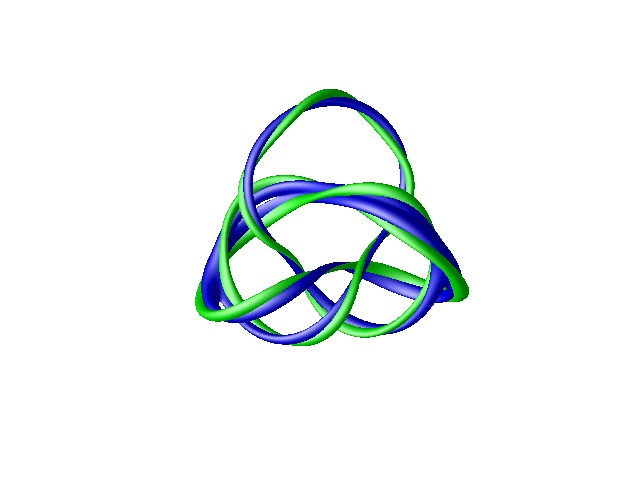}
	\caption{$22\mathcal{L}_{9(3,2), 5}^{4,4}$}
	\label{fig::22l95}
\end{subfigure}
\begin{subfigure}[b]{0.19\textwidth}
	\includegraphics[width=\textwidth, clip=true, trim= 35 0 35 0]{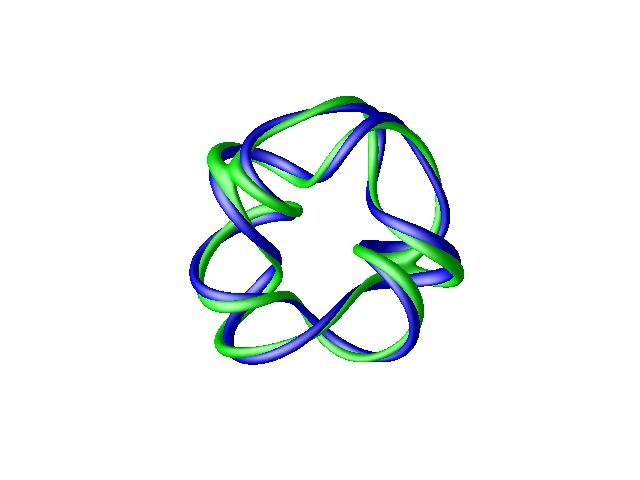}
	\caption{$22\mathcal{K}_{7,2}$}
	\label{fig::22k72}
\end{subfigure}
\begin{subfigure}[b]{0.19\textwidth}
	\includegraphics[width=\textwidth, clip=true, trim= 35 0 35 0]{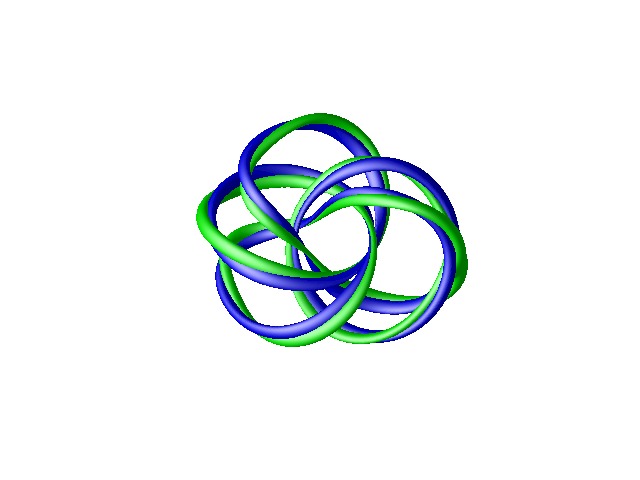}
	\caption{$23\mathcal{K}_{5,4}$}
	\label{fig::23k54}
\end{subfigure}
\begin{subfigure}[b]{0.19\textwidth}
	\includegraphics[width=\textwidth, clip=true, trim= 35 0 35 0]{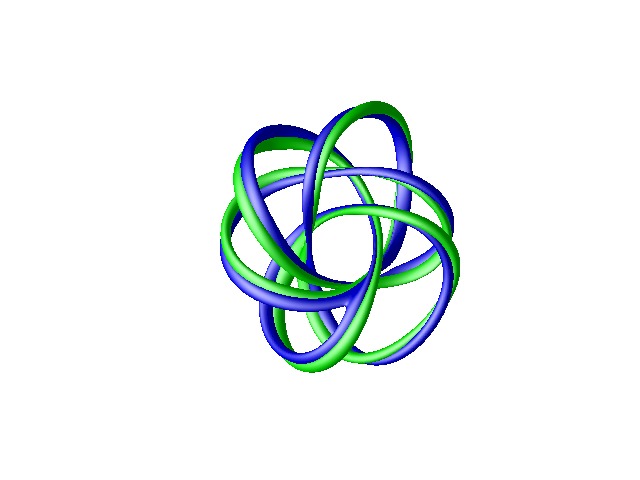}
	\caption{$\boldsymbol{23\mathcal{C}^{2,5}_{3,2}}$}
	\label{fig::23c52_32}
\end{subfigure}
\begin{subfigure}[b]{0.19\textwidth}
	\includegraphics[width=\textwidth, clip=true, trim= 35 0 35 0]{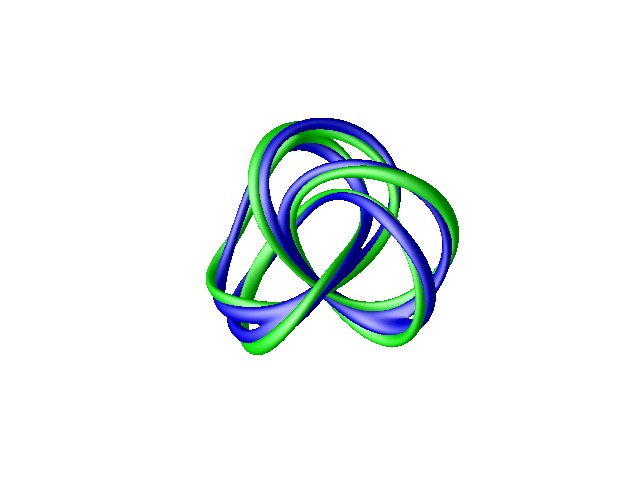}
	\caption{$23\mathcal{L}_{8(3,2),3,2}^{4,3,3}$}
	\label{fig::23l832}
\end{subfigure}
\begin{subfigure}[b]{0.19\textwidth}
	\includegraphics[width=\textwidth, clip=true, trim= 35 0 35 0]{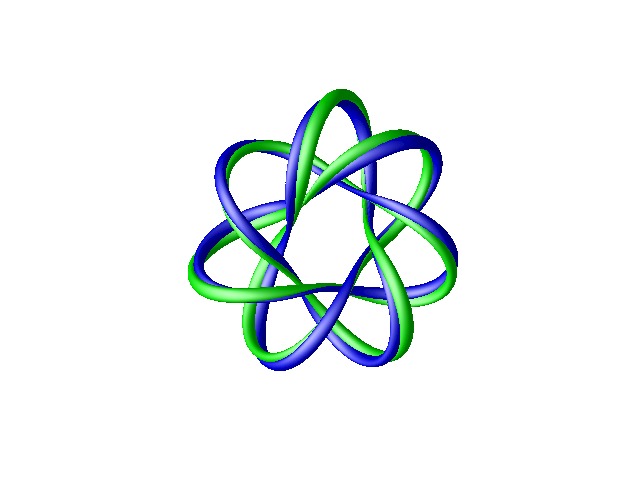}
	\caption{$23\mathcal{K}_{7,3}$}
	\label{fig::23k73}
\end{subfigure}
\begin{subfigure}[b]{0.19\textwidth}
	\includegraphics[width=\textwidth, clip=true, trim= 35 0 35 0]{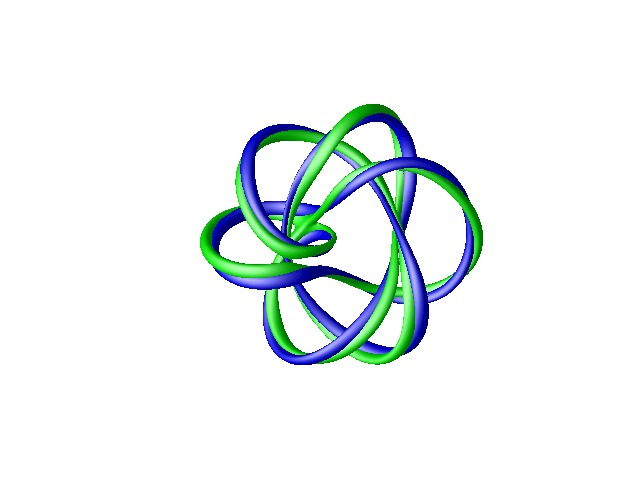}
	\caption{$23\mathcal{L}_{11(5,2), 4}^{4,4}$}
	\label{fig::23l114}
\end{subfigure}
\begin{subfigure}[b]{0.19\textwidth}
	\includegraphics[width=\textwidth, clip=true, trim= 35 0 35 0]{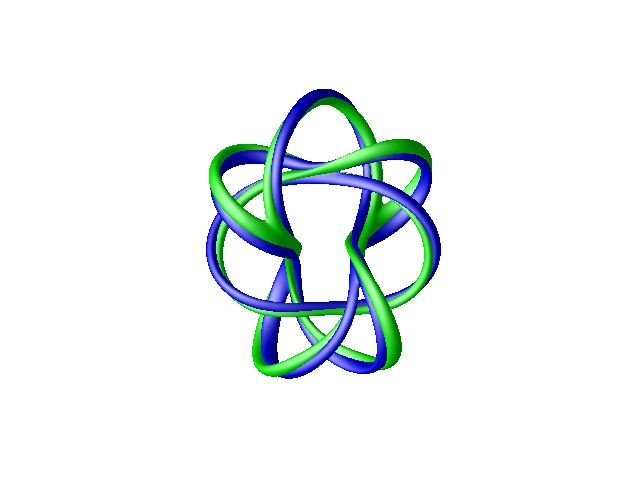}
	\caption{$23\mathcal{L}_{13(5,2), 2}^{4,4}$}
	\label{fig::23l132}
\end{subfigure}
\caption{The position curves (blue) and linking curve (green) for a range of solutions with topological charge $20\le Q \le 23$. Solutions not formed of torus knots are marked in bold.}
\label{fig::mainresult2} 
\end{figure}

\begin{figure}
\begin{subfigure}[b]{0.19\textwidth}
	\includegraphics[width=\textwidth, clip=true, trim= 35 0 35 0]{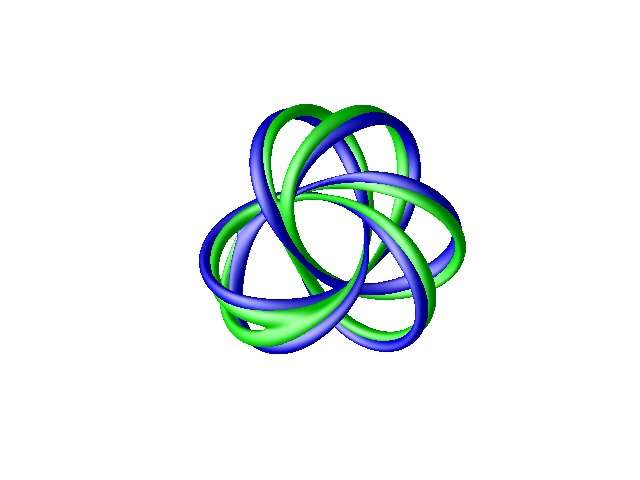}
	\caption{$24\mathcal{L}_{6(3,2),6(3,2)}^{6,6}$}
	\label{fig::24c62_32}
\end{subfigure}
\begin{subfigure}[b]{0.19\textwidth}
	\includegraphics[width=\textwidth, clip=true, trim= 35 0 35 0]{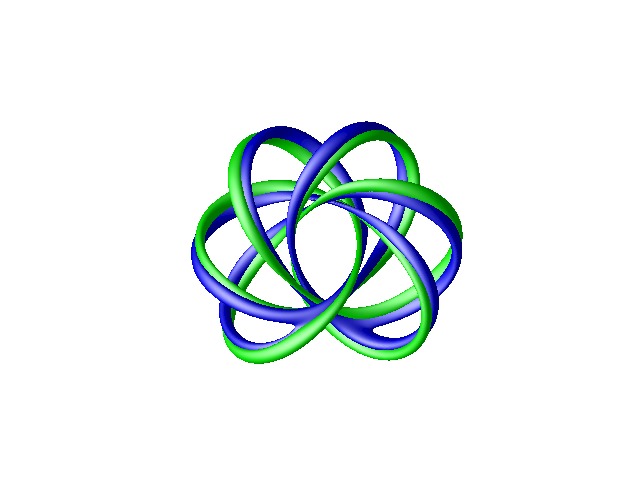}
	\caption{$24\mathcal{L}_{7(3,2),7(3,2)}^{5,5}$}
	\label{fig::24c42_32}
\end{subfigure}
\begin{subfigure}[b]{0.19\textwidth}
	\includegraphics[width=\textwidth, clip=true, trim= 35 0 35 0]{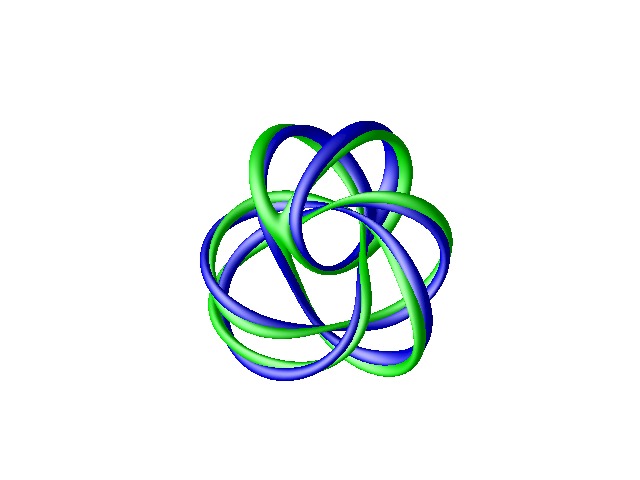}
	\caption{$24\mathcal{L}_{17(5,3), 1}^{3,3}$}
	\label{fig::24l171}
\end{subfigure}
\begin{subfigure}[b]{0.19\textwidth}
	\includegraphics[width=\textwidth, clip=true, trim= 35 0 35 0]{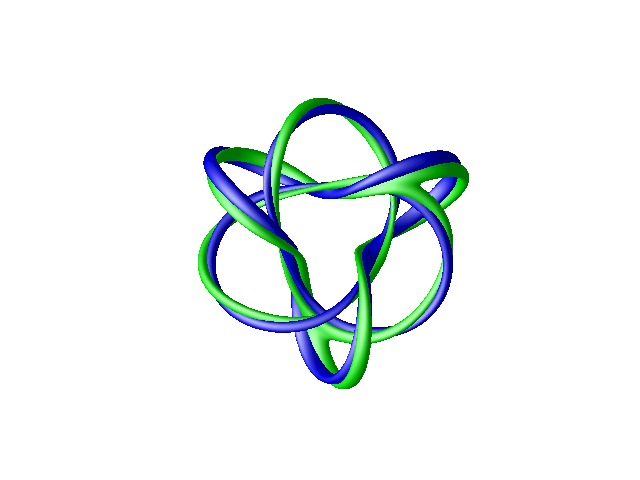}
	\caption{$24\mathcal{L}_{8(3,2), 4}^{6,6}$}
	\label{fig::24l84}
\end{subfigure}
\begin{subfigure}[b]{0.19\textwidth}
	\includegraphics[width=\textwidth, clip=true, trim= 35 0 35 0]{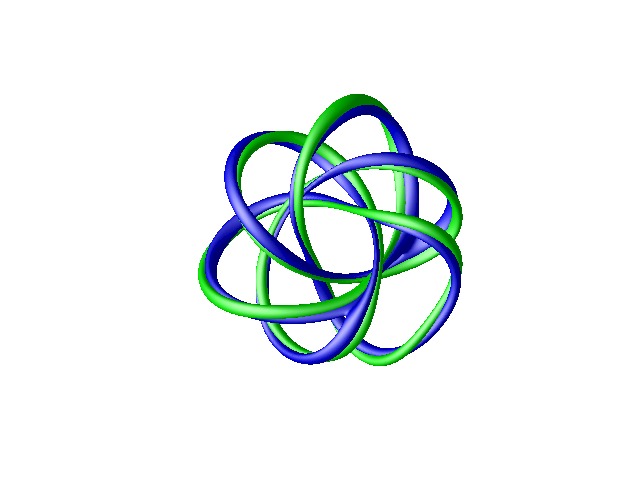}
	\caption{$25\mathcal{L}_{7(3,2),6(3,2)}^{6,6}$}
	\label{fig::25l76}
\end{subfigure}
\begin{subfigure}[b]{0.19\textwidth}
	\includegraphics[width=\textwidth, clip=true, trim= 35 0 35 0]{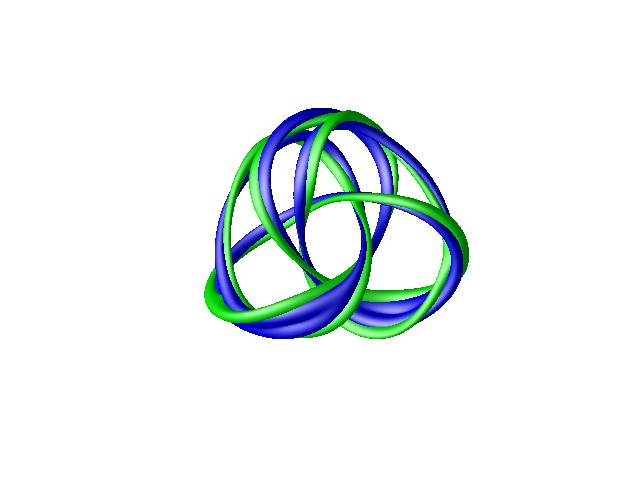}
	\caption{$25\mathcal{K}_{5,4}$}
	\label{fig::25k54}
\end{subfigure}
\begin{subfigure}[b]{0.19\textwidth}
	\includegraphics[width=\textwidth, clip=true, trim= 35 0 35 0]{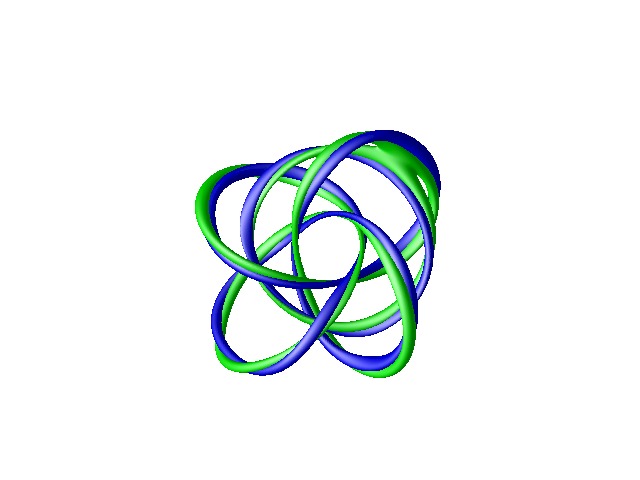}
	\caption{$25\mathcal{L}_{15(4,3),2}^{4,4}$}
	\label{fig::25l152}
\end{subfigure}
\begin{subfigure}[b]{0.19\textwidth}
	\includegraphics[width=\textwidth, clip=true, trim= 35 0 35 0]{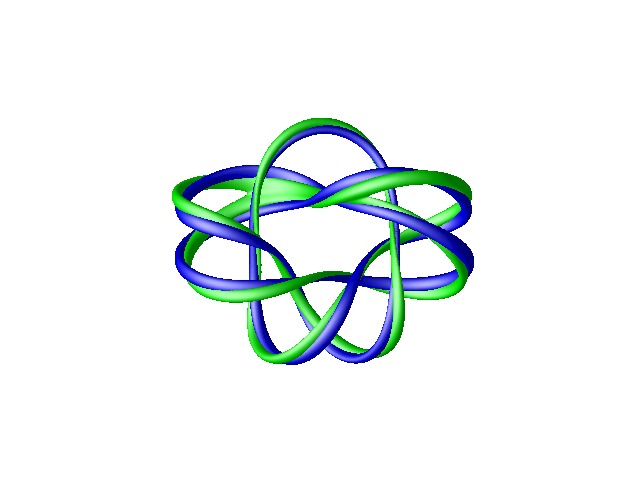}
	\caption{$25\mathcal{K}_{7,3}$}
	\label{fig::25k73}
\end{subfigure}
\begin{subfigure}[b]{0.19\textwidth}
	\includegraphics[width=\textwidth, clip=true, trim= 35 0 35 0]{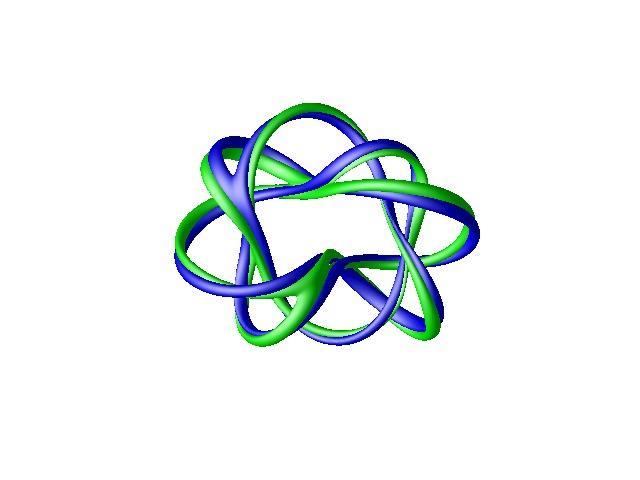}
	\caption{$25\mathcal{L}_{11(5,2),4}^{5,5}$}
	\label{fig::25l114}
\end{subfigure}
\begin{subfigure}[b]{0.19\textwidth}
	\includegraphics[width=\textwidth, clip=true, trim= 35 0 35 0]{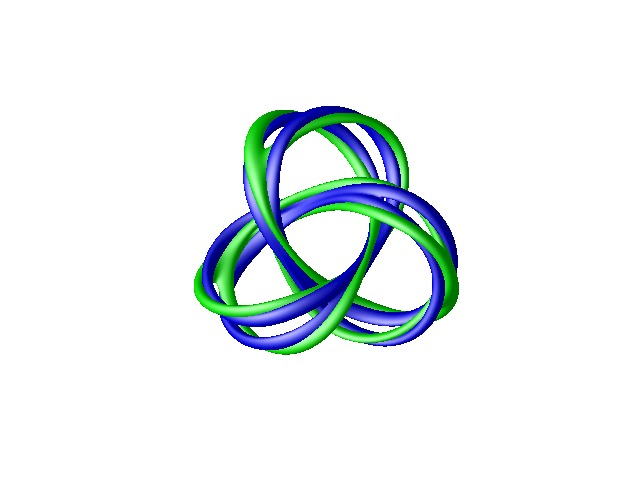}
	\caption{$26\mathcal{L}_{7(3,2),7(3,2)}^{6,6}$}
	\label{fig::26l77}
\end{subfigure}
\begin{subfigure}[b]{0.19\textwidth}
	\includegraphics[width=\textwidth, clip=true, trim= 35 0 35 0]{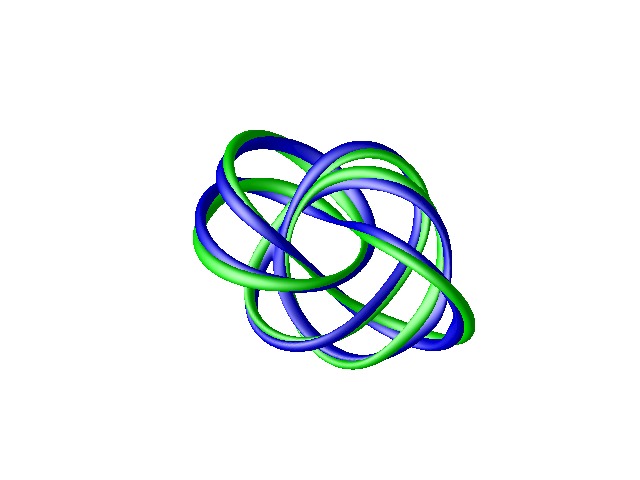}
	\caption{$26\mathcal{L}_{8(3,2),6(3,2)}^{6,6}$}
	\label{fig::26l86}
\end{subfigure}
\begin{subfigure}[b]{0.19\textwidth}
	\includegraphics[width=\textwidth, clip=true, trim= 35 0 35 0]{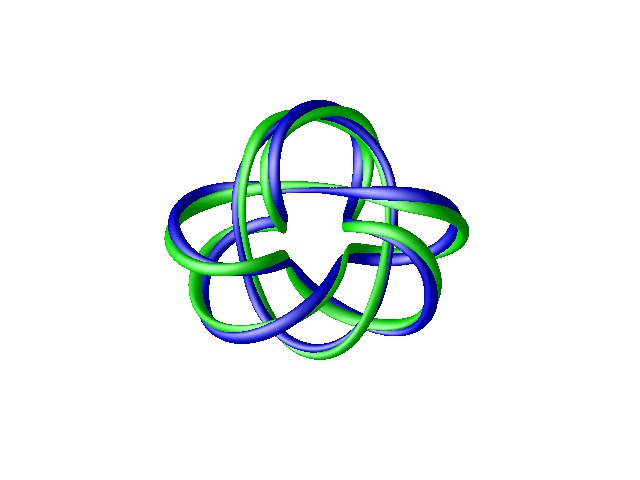}
	\caption{$26\mathcal{K}_{7,3}$}
	\label{fig::26k73}
\end{subfigure}
\begin{subfigure}[b]{0.19\textwidth}
	\includegraphics[width=\textwidth, clip=true, trim= 35 0 35 0]{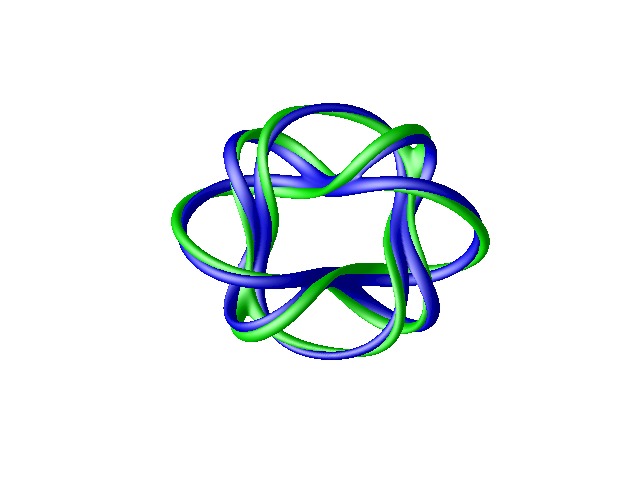}
	\caption{$26\mathcal{L}_{6,4,4}^{4,4,4}$}
	\label{fig::26l644}
\end{subfigure}
\begin{subfigure}[b]{0.19\textwidth}
	\includegraphics[width=\textwidth, clip=true, trim= 35 0 35 0]{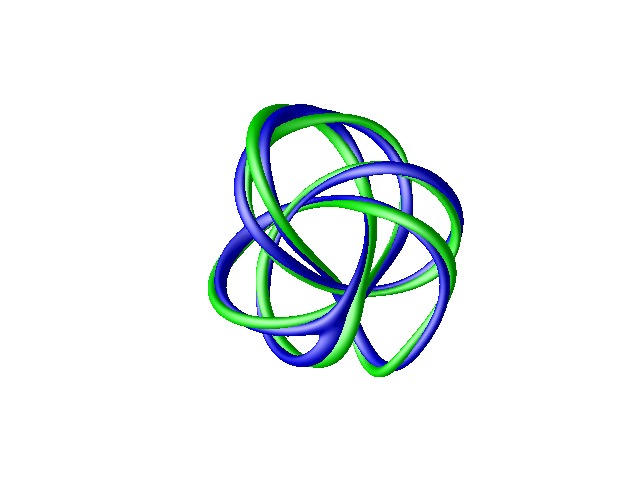}
	\caption{$\boldsymbol{27\mathcal{C}_{3,2}^{2,5}}$}
	\label{fig::27c52}
\end{subfigure}
\begin{subfigure}[b]{0.19\textwidth}
	\includegraphics[width=\textwidth, clip=true, trim= 35 0 35 0]{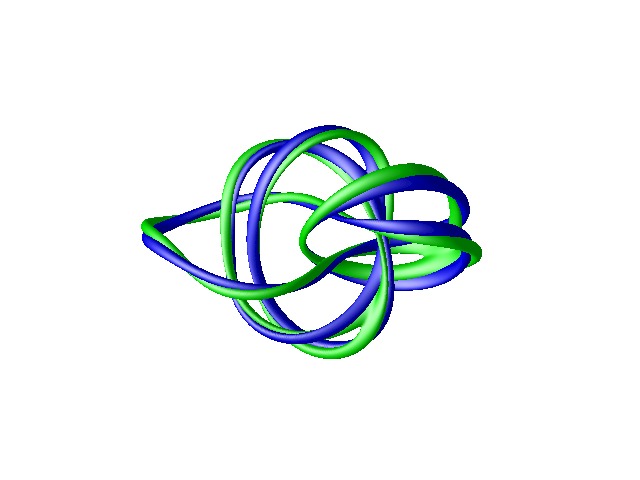}
	\caption{$27\mathcal{L}_{9(3,2),6(3,2)}^{6,6}$}
	\label{fig::27l96}
\end{subfigure}
\begin{subfigure}[b]{0.19\textwidth}
	\includegraphics[width=\textwidth, clip=true, trim= 35 0 35 0]{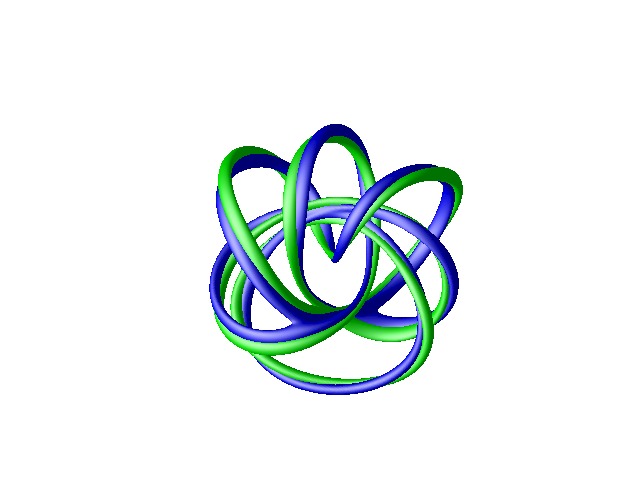}
	\caption{$27\mathcal{L}_{6(3,2),3,2}^{6,6,4}$}
	\label{fig::27l632}
\end{subfigure}
\begin{subfigure}[b]{0.19\textwidth}
	\includegraphics[width=\textwidth, clip=true, trim= 35 0 35 0]{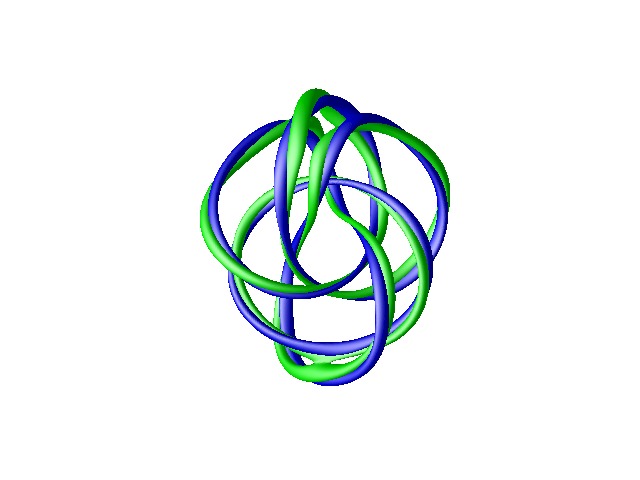}
	\caption{$27\mathcal{L}_{17(4,3),2}^{4,4}$}
	\label{fig::27l172}
\end{subfigure}
\begin{subfigure}[b]{0.19\textwidth}
	\includegraphics[width=\textwidth, clip=true, trim= 35 0 35 0]{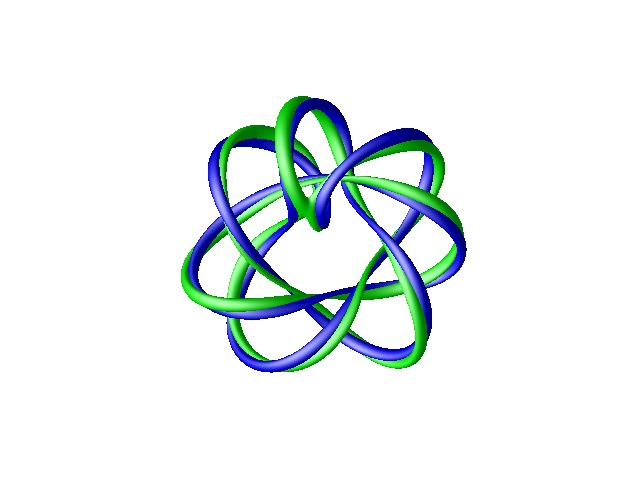}
	\caption{$27\mathcal{L}_{14(5,2),3}^{5,5}$}
	\label{fig::27l143}
\end{subfigure}
\begin{subfigure}[b]{0.19\textwidth}
	\includegraphics[width=\textwidth, clip=true, trim= 35 0 35 0]{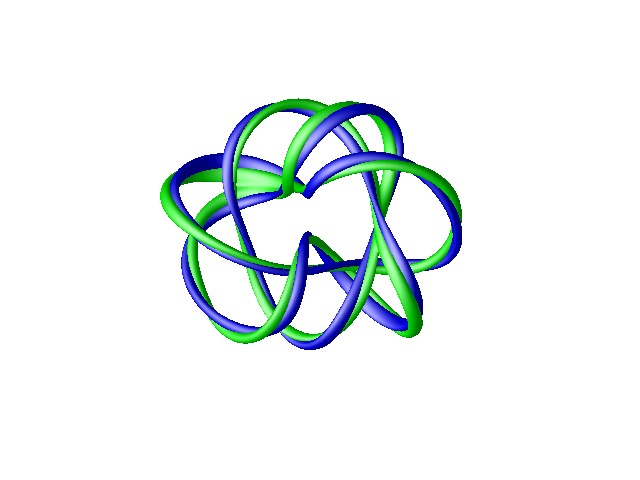}
	\caption{$27\mathcal{L}_{13(5,2),4}^{5,5}$}
	\label{fig::27l134}
\end{subfigure}
\begin{subfigure}[b]{0.19\textwidth}
	\includegraphics[width=\textwidth, clip=true, trim= 35 0 35 0]{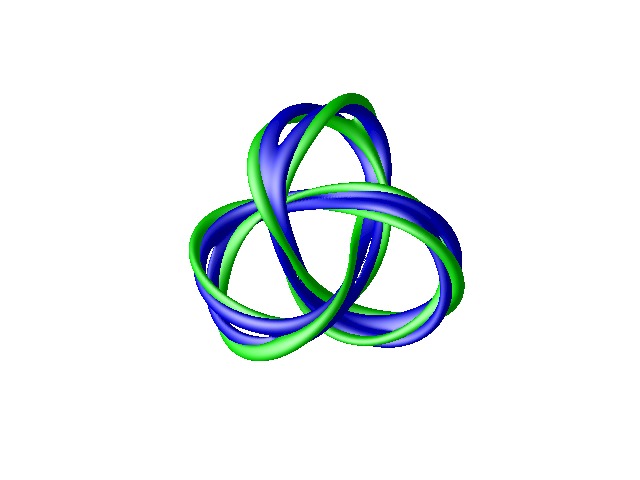}
	\caption{$\boldsymbol{28\mathcal{C}_{3,2}^{2,3}}$}
	\label{fig::28c32}
\end{subfigure}
\begin{subfigure}[b]{0.19\textwidth}
	\includegraphics[width=\textwidth, clip=true, trim= 35 0 35 0]{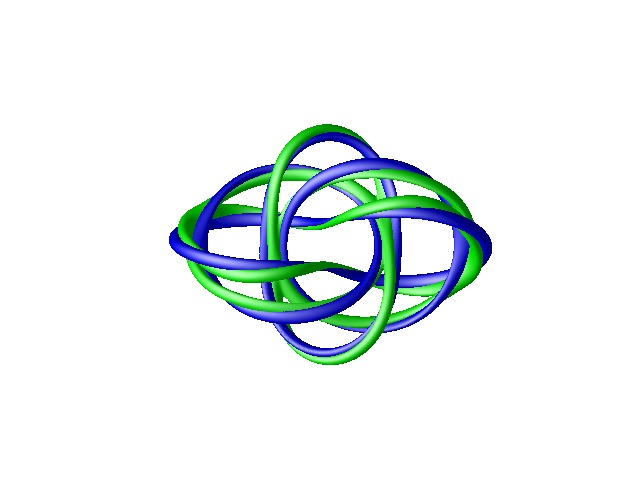}
	\caption{$28\mathcal{L}_{4,2,2,2}^{6,4,4,4}$}
	\label{fig::28l4222}
\end{subfigure}
\begin{subfigure}[b]{0.19\textwidth}
	\includegraphics[width=\textwidth, clip=true, trim= 35 0 35 0]{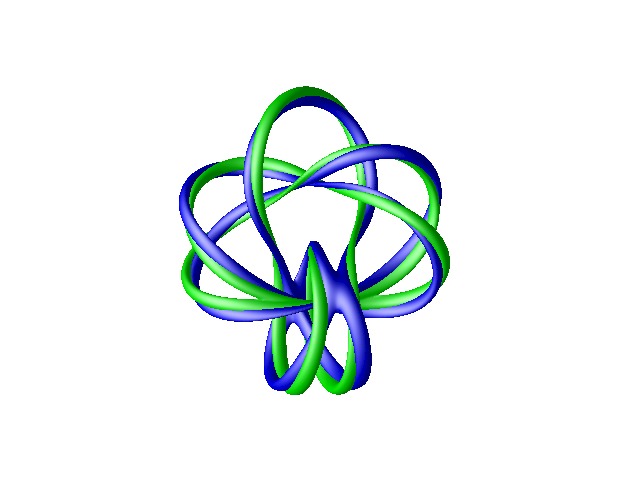}
	\caption{$28\mathcal{L}_{6,3,3}^{6,5,5}$}
	\label{fig::28l633}
\end{subfigure}
\begin{subfigure}[b]{0.19\textwidth}
	\includegraphics[width=\textwidth, clip=true, trim= 35 0 35 0]{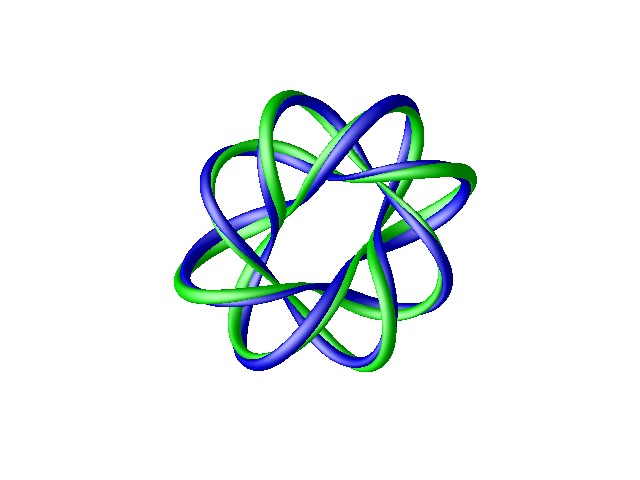}
	\caption{$28\mathcal{K}_{8,3}$}
	\label{fig::28k83}
\end{subfigure}
\begin{subfigure}[b]{0.19\textwidth}
	\includegraphics[width=\textwidth, clip=true, trim= 35 0 35 0]{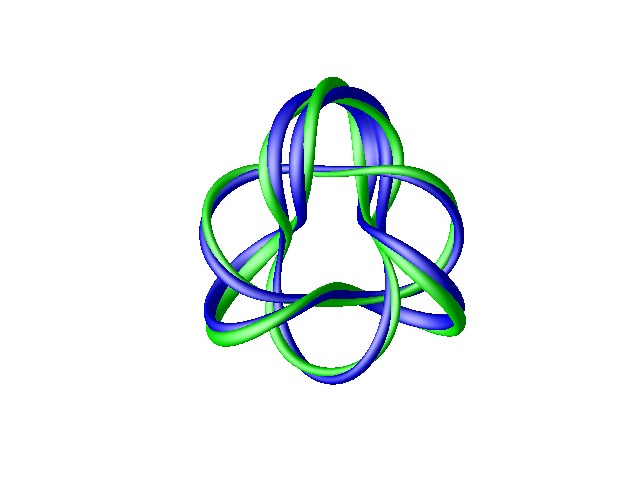}
	\caption{$28\mathcal{K}_{7,3}$}
	\label{fig::28k73}
\end{subfigure}
\caption{The position curves (blue) and linking curve (green) for a range of solutions with topological charge $24\le Q \le 28$. Solutions not formed of torus knots are marked in bold.}
\label{fig::mainresult3} 
\end{figure}

\begin{figure}
\begin{subfigure}[b]{0.19\textwidth}
	\includegraphics[width=\textwidth, clip=true, trim= 35 0 35 0]{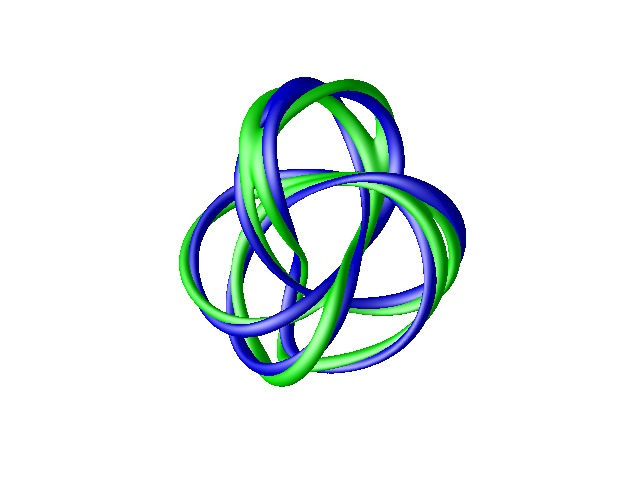}
	\caption{$29\mathcal{L}_{9(3,2),8(3,2)}^{6,6}$}
	\label{fig::29l98}
\end{subfigure}
\begin{subfigure}[b]{0.19\textwidth}
	\includegraphics[width=\textwidth, clip=true, trim= 35 0 35 0]{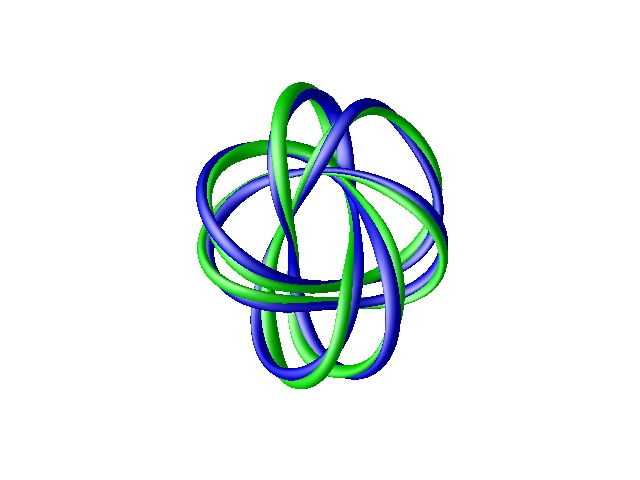}
	\caption{$29\mathcal{L}_{17(5,3),2}^{5,5}$}
	\label{fig::29l172}
\end{subfigure}
\begin{subfigure}[b]{0.19\textwidth}
	\includegraphics[width=\textwidth, clip=true, trim= 35 0 35 0]{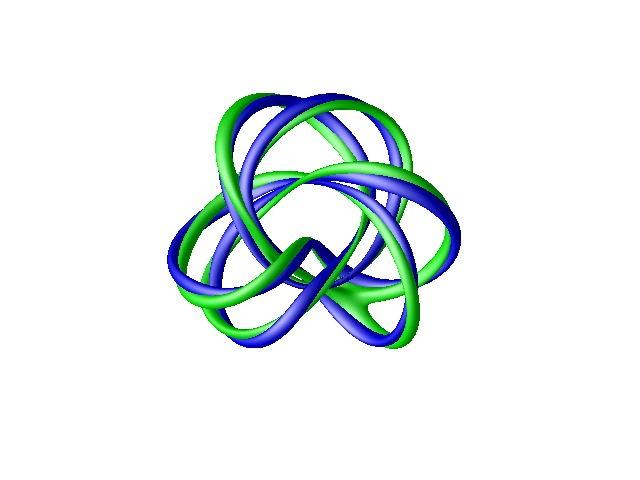}
	\caption{$\boldsymbol{29\mathcal{C}_{3,2}^{2,7}}$}
	\label{fig::29c72}
\end{subfigure}
\begin{subfigure}[b]{0.19\textwidth}
	\includegraphics[width=\textwidth, clip=true, trim= 35 0 35 0]{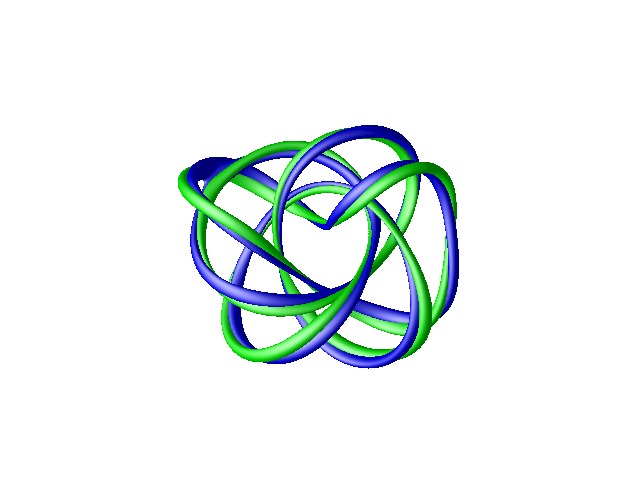}
	\caption{$29\mathcal{L}_{13(4,3),4}^{6,6}$}
	\label{fig::29l134}
\end{subfigure}
\begin{subfigure}[b]{0.19\textwidth}
	\includegraphics[width=\textwidth, clip=true, trim= 35 0 35 0]{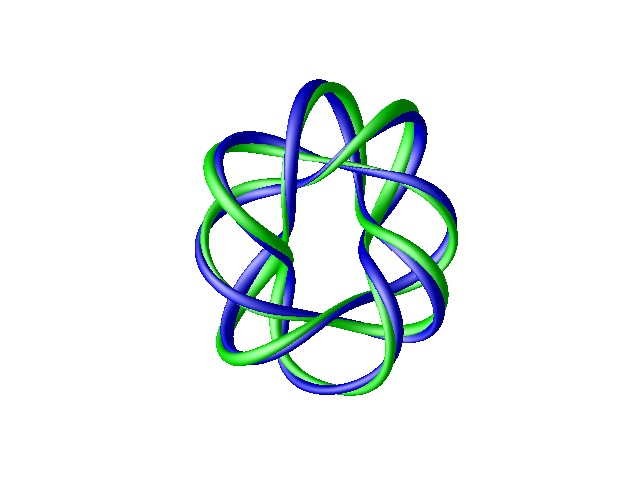}
	\caption{$29\mathcal{K}_{8,3}$}
	\label{fig::29k83}
\end{subfigure}
\begin{subfigure}[b]{0.19\textwidth}
	\includegraphics[width=\textwidth, clip=true, trim= 35 0 35 0]{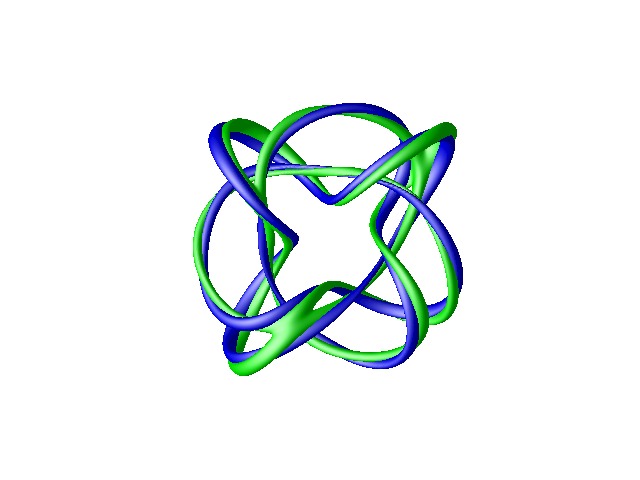}
	\caption{$29\mathcal{L}_{5,4,4}^{6,5,5}$}
	\label{fig::29l544}
\end{subfigure}
\begin{subfigure}[b]{0.19\textwidth}
	\includegraphics[width=\textwidth, clip=true, trim= 35 0 35 0]{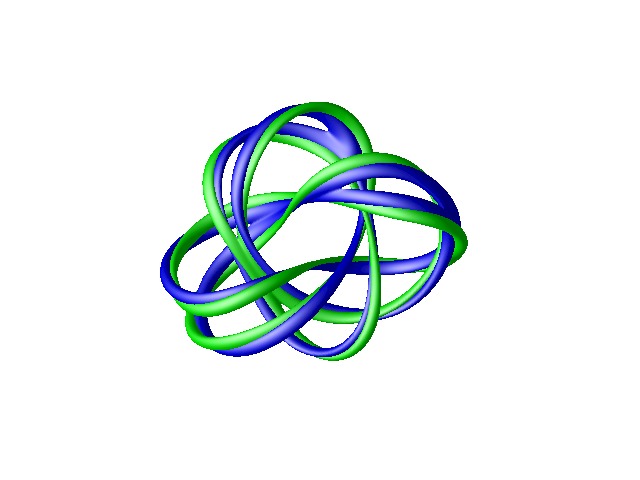}
	\caption{$30\mathcal{L}_{18(5,3),2}^{5,5}$}
	\label{fig::30l182}
\end{subfigure}
\begin{subfigure}[b]{0.19\textwidth}
	\includegraphics[width=\textwidth, clip=true, trim= 35 0 35 0]{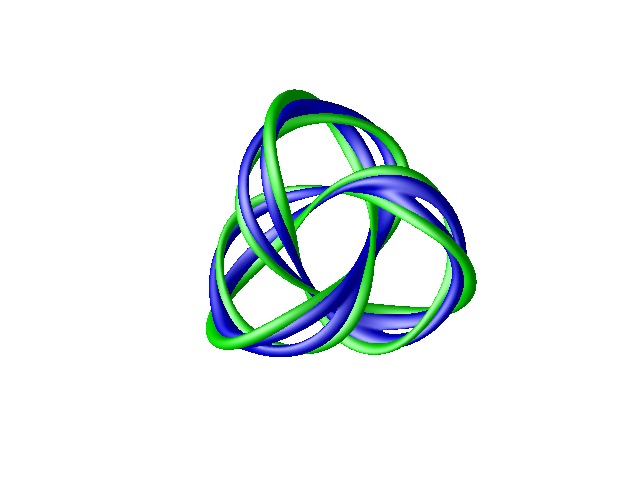}
	\caption{$\boldsymbol{30\mathcal{C}_{3,2}^{2,5}}$}
	\label{fig::30c52}
\end{subfigure}
\begin{subfigure}[b]{0.19\textwidth}
	\includegraphics[width=\textwidth, clip=true, trim= 35 0 35 0]{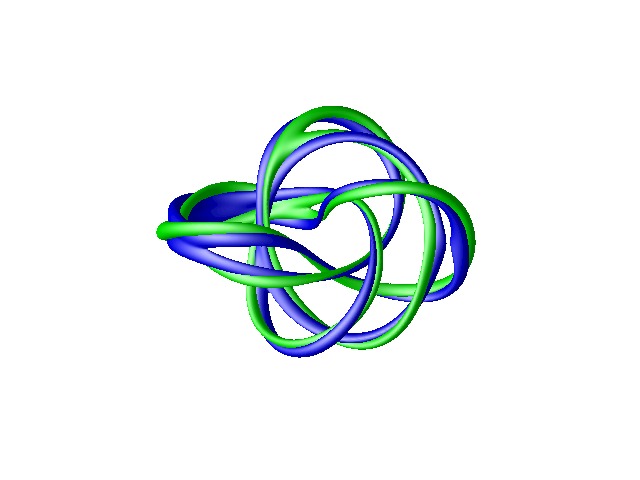}
	\caption{$\boldsymbol{30\mathcal{C}_{3,2}^{2,7}}$}
	\label{fig::30c72}
\end{subfigure}
\begin{subfigure}[b]{0.19\textwidth}
	\includegraphics[width=\textwidth, clip=true, trim= 35 0 35 0]{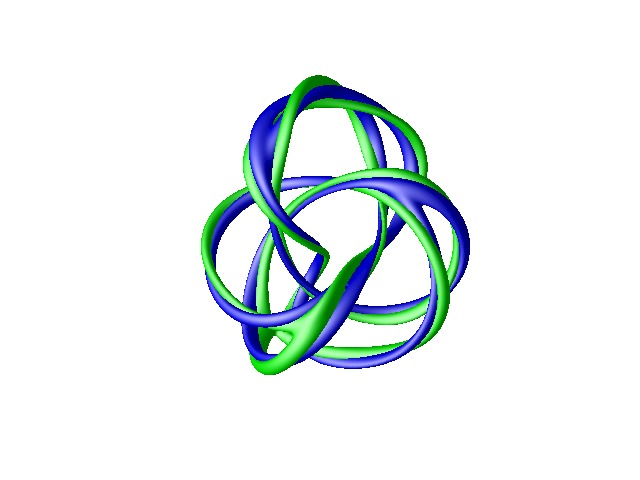}
	\caption{$\boldsymbol{31\mathcal{C}_{3,2}^{2,7}}$}
	\label{fig::31c72}
\end{subfigure}
\begin{subfigure}[b]{0.19\textwidth}
	\includegraphics[width=\textwidth, clip=true, trim= 35 0 35 0]{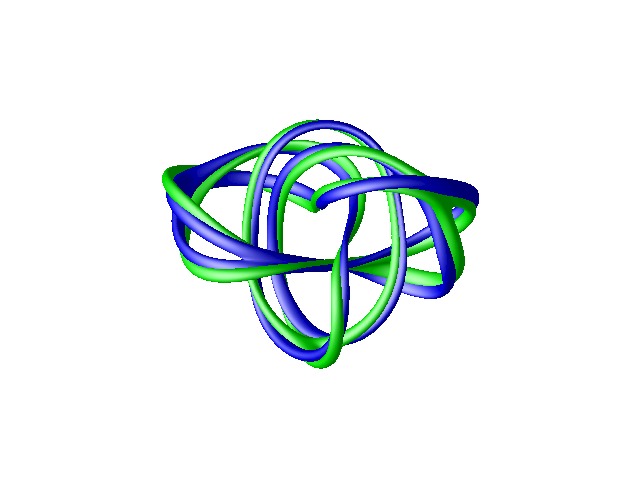}
	\caption{$31\mathcal{L}_{15(4,3),4}^{6,6}$}
	\label{fig::31l154}
\end{subfigure}
\begin{subfigure}[b]{0.19\textwidth}
	\includegraphics[width=\textwidth, clip=true, trim= 35 0 35 0]{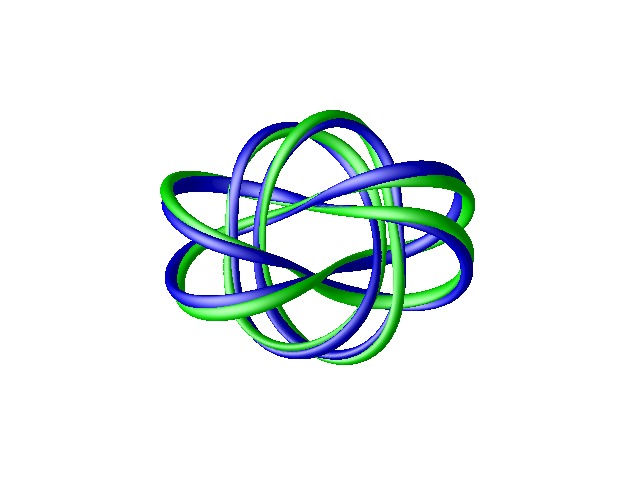}
	\caption{$32\mathcal{L}_{3,3,2,2}^{6,6,5,5}$}
	\label{fig::32l3322}
\end{subfigure}
\begin{subfigure}[b]{0.19\textwidth}
	\includegraphics[width=\textwidth, clip=true, trim= 35 0 35 0]{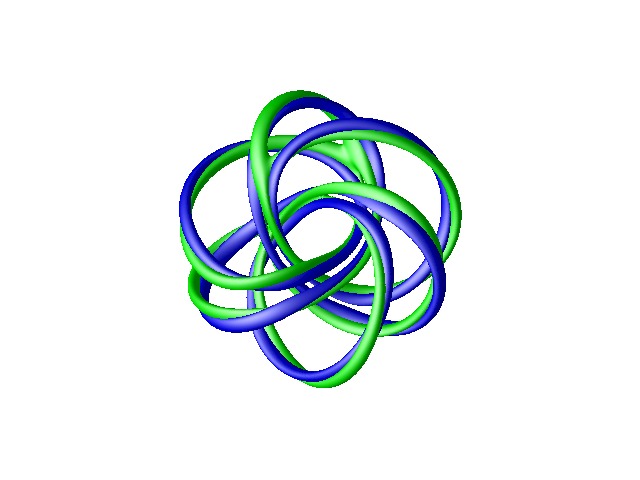}
	\caption{$32\mathcal{L}_{8(3,2),2,2,2}^{6,4,4,4}$}
	\label{fig::32l8222}
\end{subfigure}
\begin{subfigure}[b]{0.19\textwidth}
	\includegraphics[width=\textwidth, clip=true, trim= 35 0 35 0]{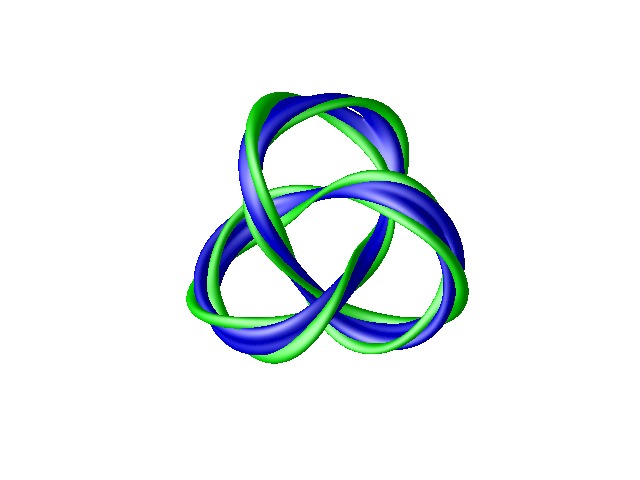}
	\caption{$\boldsymbol{32\mathcal{C}_{3,2}^{2,5}}$}
	\label{fig::32c52}
\end{subfigure}
\begin{subfigure}[b]{0.19\textwidth}
	\includegraphics[width=\textwidth, clip=true, trim= 35 0 35 0]{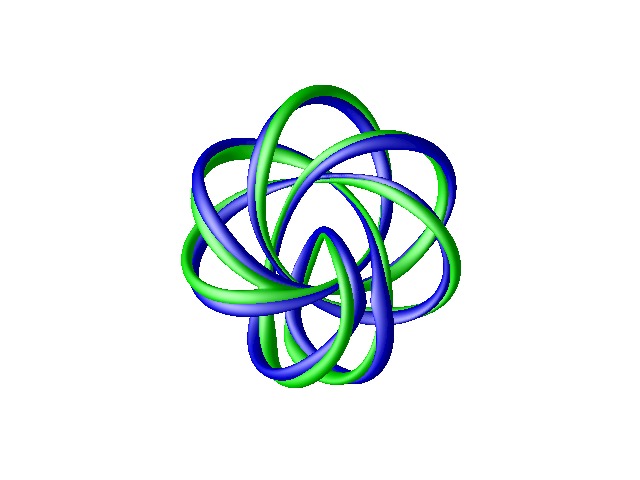}
	\caption{$33\mathcal{L}_{11(5,2),6(3,2)}^{8,8}$}
	\label{fig::33l116}
\end{subfigure}
\begin{subfigure}[b]{0.19\textwidth}
	\includegraphics[width=\textwidth, clip=true, trim= 35 0 35 0]{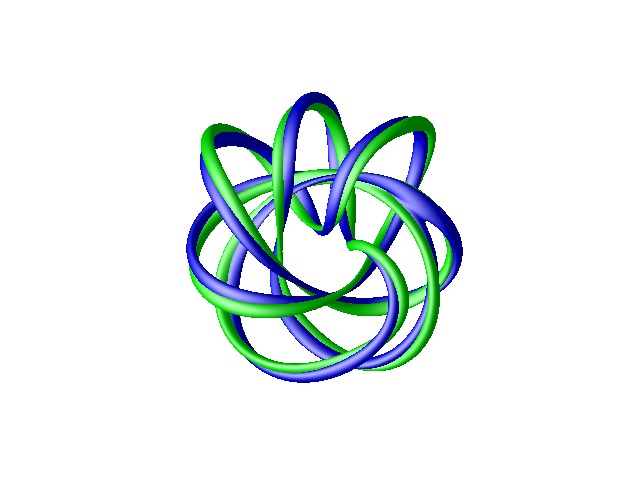}
	\caption{$\boldsymbol{33\mathcal{H}_{3.609}}$}
	\label{fig::33h}
\end{subfigure}
\begin{subfigure}[b]{0.19\textwidth}
	\includegraphics[width=\textwidth, clip=true, trim= 35 0 35 0]{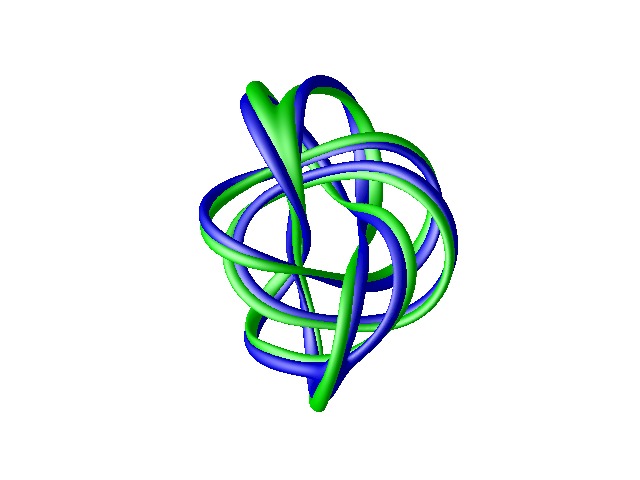}
	\caption{$33\mathcal{L}_{3,3,2,1}^{7,7,5,5}$}
	\label{fig::33l3321}
\end{subfigure}
\caption{The position curves (blue) and linking curve (green) for a range of solutions with topological charge $29\le Q \le 33$. Solutions not formed of torus knots are marked in bold.}
\label{fig::mainresult4} 
\end{figure}

\begin{figure}

\begin{subfigure}[b]{0.19\textwidth}
	\includegraphics[width=\textwidth, clip=true, trim= 35 0 35 0]{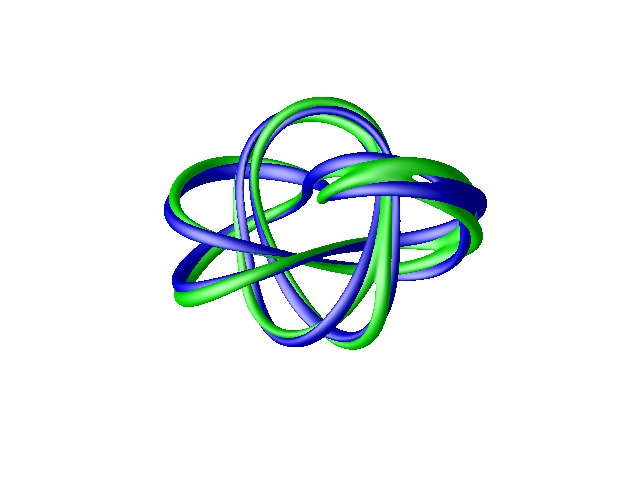}
	\caption{$34\mathcal{L}_{4,4,2,2}^{6,6,5,5}$}
	\label{fig::34l4422}
\end{subfigure}
\begin{subfigure}[b]{0.19\textwidth}
	\includegraphics[width=\textwidth, clip=true, trim= 35 0 35 0]{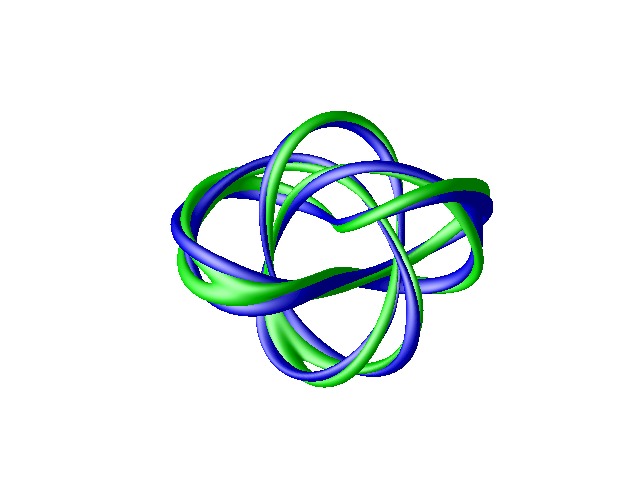}
	\caption{$\boldsymbol{34\mathcal{H}_{3.609}}$}
	\label{fig::34h}
\end{subfigure}
\begin{subfigure}[b]{0.19\textwidth}
	\includegraphics[width=\textwidth, clip=true, trim= 35 0 35 0]{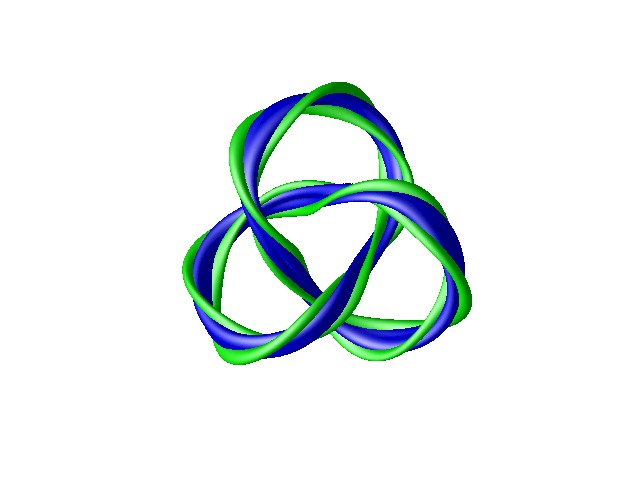}
	\caption{$\boldsymbol{34\mathcal{C}_{3,2}^{2,3}}$}
	\label{fig::34c32}
\end{subfigure}
\begin{subfigure}[b]{0.19\textwidth}
	\includegraphics[width=\textwidth, clip=true, trim= 35 0 35 0]{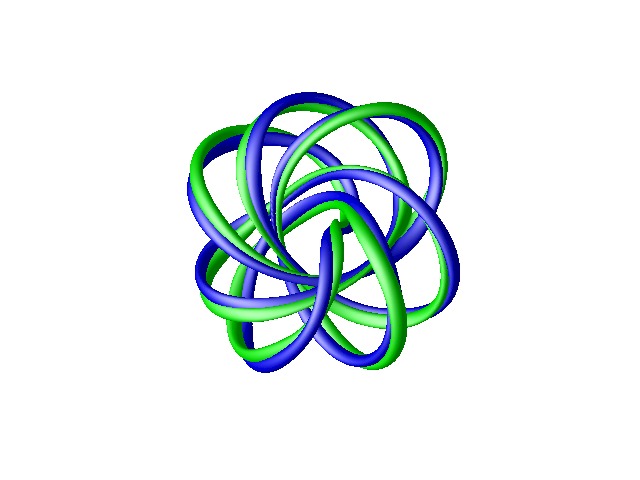}
	\caption{$35\mathcal{L}_{13(4,3),6(3,2)}^{8,8}$}
	\label{fig::35l136}
\end{subfigure}
\begin{subfigure}[b]{0.19\textwidth}
	\includegraphics[width=\textwidth, clip=true, trim= 35 0 35 0]{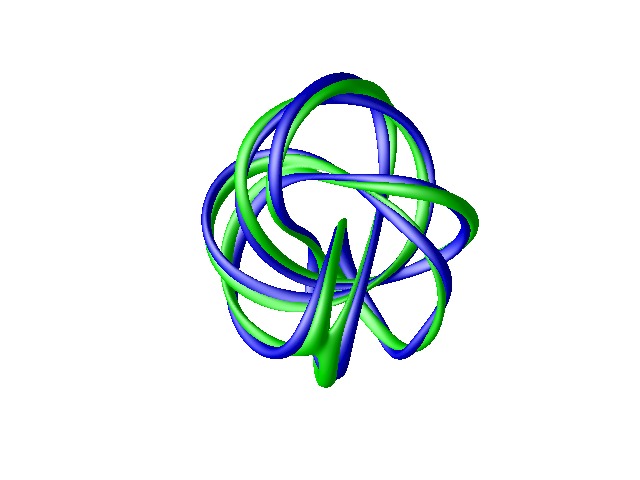}
	\caption{$\boldsymbol{35\mathcal{H}_{3.609}}$}
	\label{fig::35h}
\end{subfigure}
\begin{subfigure}[b]{0.19\textwidth}
	\includegraphics[width=\textwidth, clip=true, trim= 35 0 35 0]{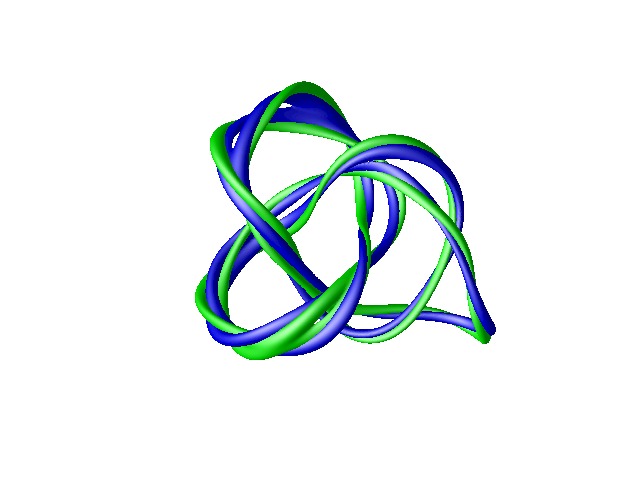}
	\caption{$35\mathcal{L}_{11(3,2),10(3,2)}^{7,7}$}
	\label{fig::35l1110}
\end{subfigure}
\begin{subfigure}[b]{0.19\textwidth}
	\includegraphics[width=\textwidth, clip=true, trim= 35 0 35 0]{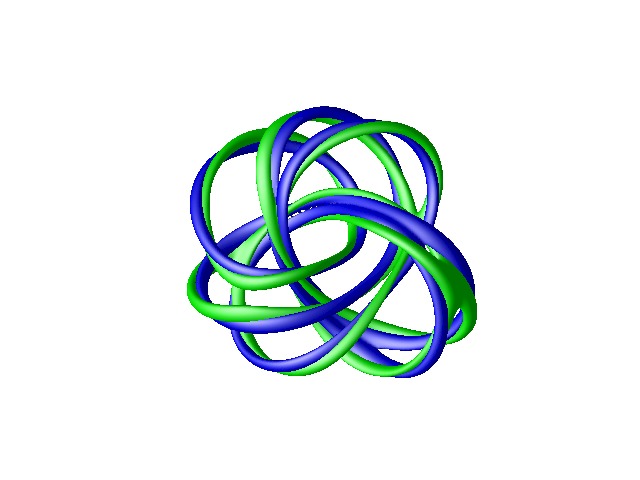}
	\caption{$\boldsymbol{36\mathcal{L}_{26(2,3;3,2),2}^{4,4}}$}
	\label{fig::36l262}
\end{subfigure}
\begin{subfigure}[b]{0.19\textwidth}
	\includegraphics[width=\textwidth, clip=true, trim= 35 0 35 0]{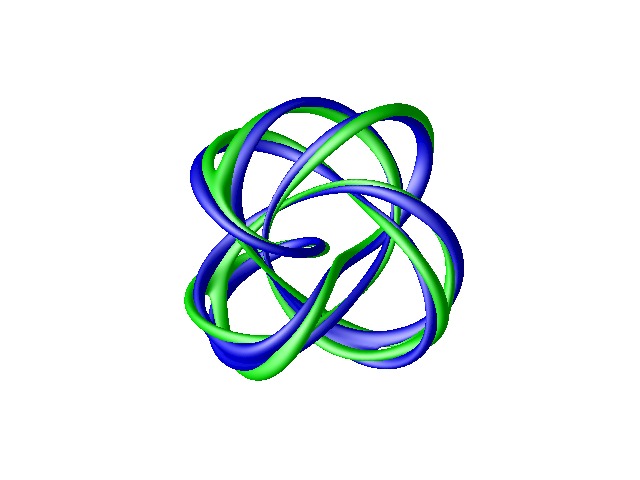}
	\caption{$\boldsymbol{36\mathcal{H}_{3.609}}$}
	\label{fig::36h}
\end{subfigure}
\begin{subfigure}[b]{0.19\textwidth}
	\includegraphics[width=\textwidth, clip=true, trim= 35 0 35 0]{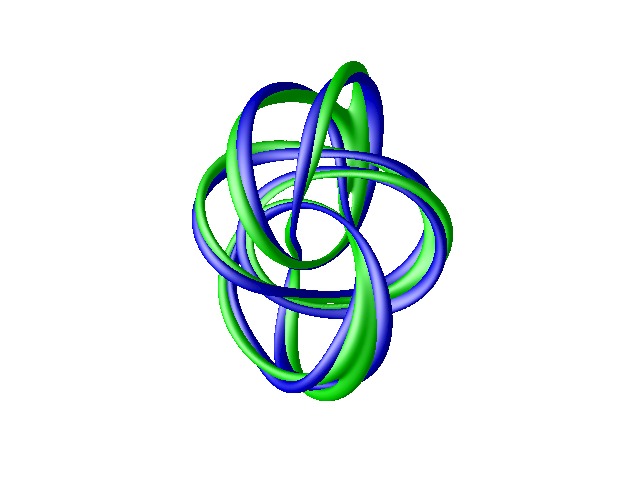}
\caption{$36\mathcal{L}_{15(4,3),2,1}^{8,5,5}$}
	\label{fig::36l1521}
\end{subfigure}
\begin{subfigure}[b]{0.19\textwidth}
	\includegraphics[width=\textwidth, clip=true, trim= 35 0 35 0]{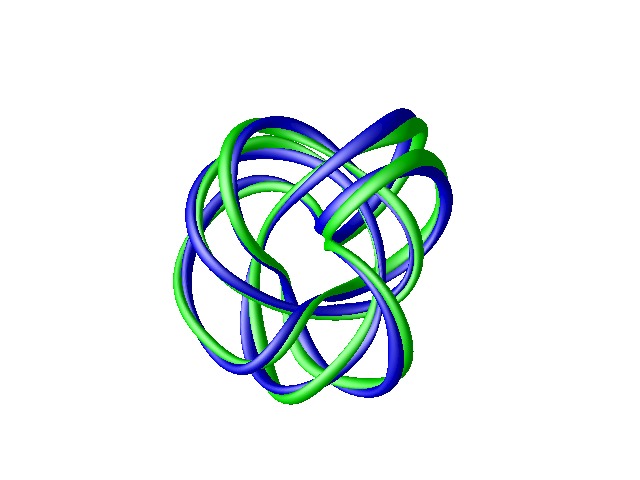}
\caption{$\boldsymbol{36\mathcal{L}_{20\mathcal{H}_{2.828},4}^{6,6}}$}
	\label{fig::36l204}
\end{subfigure}
\caption{The position curves (blue) and linking curve (green) for a range of solutions with topological charge $34\le Q \le 36$. Solutions not formed of torus knots are marked in bold.}
\label{fig::mainresult5} 
\end{figure}

We now present the results for hopfions with topological charge sixteen to thirty--six. For each topological charge a range of initial field configurations are subjected to the energy minimisation scheme, and we record in Tables \ref{tab::energies1} and \ref{tab::energies2} the energy of configurations, along with a comparison to Ward's conjectured bound. We find that solutions for higher charges obey a similar growth in energy as found for lower charge solutions and in the other Skyrme-type models, exceeding Ward's bound by a little over 20\%. Plots of solutions showing the location curve and curve showing twist can be found in Figures \ref{fig::mainresult1}, \ref{fig::mainresult2}, \ref{fig::mainresult3}, \ref{fig::mainresult4} and \ref{fig::mainresult5}.

\begin{table}
\footnotesize\setlength{\tabcolsep}{2.5pt}
\hspace*{-3cm}
\centering
\begin{tabular}[t]{|c|c|c|c|}
\hline
$\phantom{Q}Q\phantom{Q}$ & Initial $\rightarrow$ final & $E$ & $E/Q^{3/4}$\\
\hline\hline
\multirow{6}{*}{$16$} & \rule{0pt}{2.5ex}$\mathcal{K}_{3,2},\mathcal{K}_{5,3}\rightarrow\mathcal{K}_{5,3}$ & 9.887 & 1.236 \\
\cline{2-4}
\rule{0pt}{2.5ex}&$\mathcal{K}_{5,4}\rightarrow \mathcal{L}_{4,4}^{4,4}$ & 9.904 & 1.238 \\
\cline{2-4}
\rule{0pt}{2.5ex}& $\mathcal{L}_{4,3,3}^{2,2,2},\mathcal{C}^{2,5}_{3,2}, \mathcal{C}^{2,7}_{3,2}\rightarrow \mathcal{L}_{8(3,2),2}^{3,3}$& 9.919 & 1.240\\
\cline{2-4}
\rule{0pt}{2.5ex}&$\mathcal{C}^{2,3}_{3,2},\mathcal{K}_{4,3}\rightarrow \mathcal{K}_{4,3}$ & 9.969 & 1.246 \\
\cline{2-4}
\rule{0pt}{2.5ex} &$\mathcal{L}^{5,5}_{4(3,2) ,2(3,2)}\rightarrow\mathcal{L}_{5(3,2),3}^{4,4}$ & 10.057 & 1.257\\
\cline{2-4}
\rule{0pt}{2.5ex}&$\mathcal{K}_{5,2}\rightarrow \mathcal{L}_{7(3,2),3}^{3,3}$ & 10.165 & 1.271\\\cline{2-4}
\hline\hline
\multirow{5}{*}{$17$} &\rule{0pt}{2.5ex}$\mathcal{K}_{5,3},\mathcal{K}_{7,3}\rightarrow \mathcal{L}_{8(3,2),3}^{3,3}$ & 10.355 & 1.237  \\\cline{2-4}
\rule{0pt}{2.5ex}& $\mathcal{K}_{4,3}\rightarrow \mathcal{L}_{9(3,2),2}^{3,3}$& 10.422 & 1.245 \\ \cline{2-4}
\rule{0pt}{2.5ex} & $\mathcal{C}^{2,5}_{3,2}\rightarrow \mathcal{L}_{2,1,1,1}^{3,3,3,3}$ & 10.424 & 1.245\\\cline{2-4}
\rule{0pt}{2.5ex}& $\mathcal{C}^{2,3}_{3,2},\mathcal{C}^{2,7}_{3,2}, \mathcal{K}_{5,4}\rightarrow \mathcal{L}_{5(3,2),1,1}^{4,3,3}$ & 10.487 & 1.253 \\\cline{2-4}
\rule{0pt}{2.5ex}&$\mathcal{K}_{7,2}\rightarrow \mathcal{K}_{7,2}$ & 10.952  & 1.308\\\cline{2-4}
\hline\hline
\multirow{6}{*}{$18$}\rule{0pt}{2.5ex} &$\mathcal{K}_{4,3}, \mathcal{K}_{5,3}\rightarrow \mathcal{K}_{5,3}$ & 10.821 & 1.238 \\\cline{2-4}
\rule{0pt}{2.5ex}& $\mathcal{C}^{2,3}_{3,2}\rightarrow \mathcal{L}_{6(3,2), 1,1}^{ 4,3,3}$ & 10.832 & 1.240 \\\cline{2-4}
\rule{0pt}{2.5ex} &$\begin{matrix} \rule{0pt}{2.5ex} \mathcal{C}^{2,5}_{3,2},\mathcal{C}^{2,7}_{3,2}, \mathcal{K}_{5,4} \\ \rule{0pt}{2.5ex} \mathcal{L}^{5,5}_{4(3,2),4(3,2)}\end{matrix}\rightarrow \mathcal{L}_{11(5,2), 1}^{3,3}$ & 10.835 & 1.240 \\\cline{2-4}
\rule{0pt}{2.5ex} &$\mathcal{K}_{7,2}\rightarrow \boldsymbol{\mathcal{H}_{4.851}}$ & 10.838 & 1.240 \\ \cline{2-4}
\rule{0pt}{2.5ex} &$\mathcal{K}_{7,3}\rightarrow \mathcal{L}_{8(3,2),4}^{3,3}$ & 10.850 & 1.242 \\
\hline\hline
\multirow{6}{*}{$19$} \rule{0pt}{2.5ex}&$\mathcal{C}^{2,5}_{3,2},\mathcal{C}^{2,7}_{3,2}\rightarrow \mathcal{K}_{5,4}$ & 11.295 & 1.241 \\ \cline{2-4}
\rule{0pt}{2.5ex}& $\mathcal{K}_{4,3},\mathcal{K}_{5,3}, \mathcal{L}^{5,5}_{7(3,2),2(3,2)}$ $\rightarrow \mathcal{K}_{5,3}$ & 11.311 & 1.243 \\ \cline{2-4}
\rule{0pt}{2.5ex}
& $\mathcal{K}_{5,4}\rightarrow\mathcal{L}_{6(3,2),2,1}^{4,3,3}$ & 11.315 & 1.243\\ \cline{2-4}
\rule{0pt}{2.5ex}& $\mathcal{C}^{2,3}_{3,2}, $  $\mathcal{L}^{6,6}_{5(3,2),2(3,2)}\rightarrow \mathcal{L}_{7(3,2),4}^{4,4}$ & 11.322 & 1.245 \\ \cline{2-4}
\rule{0pt}{2.5ex}&$\mathcal{K}_{7,3}\rightarrow \mathcal{K}_{7,3}$ & 11.427 & 1.256\\ \cline{2-4}
\rule{0pt}{2.5ex}&$\mathcal{K}_{7,2}\rightarrow \mathcal{K}_{7,2}$ & 12.008 & 1.319\\ \cline{2-4}
\hline\hline
\multirow{6}{*}{$20$}\rule{0pt}{2.5ex} &$\mathcal{K}_{7,3}\rightarrow \mathcal{L}_{3,3,2}^{4,4,4}$ & 11.730 & 1.240\\ \cline{2-4}
\rule{0pt}{2.5ex}& $\mathcal{L}^{5,5}_{6(3,2),4(3,2)}\rightarrow \mathcal{K}_{5,4}$ & 11.753 & 1.243\\ \cline{2-4}
\rule{0pt}{2.5ex}& $\begin{matrix}\rule{0pt}{2.5ex}\mathcal{C}^{2,3}_{3,2},\mathcal{C}^{2,5}_{3,2},\mathcal{C}^{2,7}_{3,2},\\ \rule{0pt}{2.5ex}\mathcal{K}_{4,3}, \mathcal{L}^{6,6}_{4(3,2),4(3,2)}\end{matrix}\rightarrow \mathcal{L}^{4,4,4}_{4,2,2}$ & 11.753& 1.243 \\ \cline{2-4}
\rule{0pt}{2.5ex}&$\mathcal{K}_{5,3}\rightarrow \mathcal{L}_{9(3,2),3}^{4,4}$ & 11.832 & 1.251 \\ \cline{2-4}
\rule{0pt}{2.5ex}&$\mathcal{K}_{5,4}\rightarrow \mathcal{L}_{2,2,2,2}^{3,3,3,3}$ & 11.839 & 1.252 \\ \cline{2-4}
\rule{0pt}{2.5ex}
& $\mathcal{K}_{7,2}\rightarrow \mathcal{K}_{7,2}$ & 12.517 & 1.324\\
\hline\hline
\multirow{4}{*}{$21$} \rule{0pt}{2.5ex}& $\mathcal{C}^{2,3}_{3,2},\mathcal{C}^{2,5}_{3,2}, \mathcal{C}^{2,7}_{3,2}, \mathcal{K}_{5,4} \rightarrow \mathcal{L}_{13(4,3),2}^{3,3} $ & 12.091 & 1.233 \\ \cline{2-4}
\rule{0pt}{2.5ex}
&$\mathcal{K}_{7,3}\rightarrow \mathcal{L}_{3,3,3}^{4,4,4}$& 12.181 & 1.242\\ \cline{2-4}
\rule{0pt}{2.5ex}&  $\mathcal{L}^{6,6}_{7(3,2),2(3,2)}\rightarrow \mathcal{L}_{11(5,2),2}^{4,4}$ & 12.250 & 1.249\\ \cline{2-4}
\rule{0pt}{2.5ex}&$\mathcal{K}_{5,3}\rightarrow \mathcal{K}_{5,3}$ & 12.271 & 1.251\\ \cline{2-4}
\rule{0pt}{2.5ex}
& $\mathcal{K}_{7,2}\rightarrow \mathcal{K}_{7,2}$ & 13.075 & 1.333\\
\hline
\end{tabular}
\hspace*{1ex}
\begin{tabular}[t]{|c|c|c|c|}
\hline
$\phantom{Q}Q\phantom{Q}$ & Initial $\rightarrow$ final & $E$ & $E/Q^{3/4}$\\
\hline\hline
\multirow{7}{*}{$22$} \rule{0pt}{2.5ex}&$\mathcal{L}^{5,5}_{6(3,2),6(3,2)}\rightarrow \mathcal{L}_{8(3,2),2,2}^{4,3,3} $ & 12.446 & 1.225 \\ \cline{2-4}
\rule{0pt}{2.5ex}& $\mathcal{K}_{5,4}\rightarrow \mathcal{K}_{5,4}$ &12.455 &1.226  \\ \cline{2-4}
\rule{0pt}{2.5ex}& $\mathcal{C}^{2,3}_{3,2},\mathcal{C}^{2,5}_{3,2},\mathcal{C}^{2,7}_{3,2}\rightarrow \boldsymbol{\mathcal{C}^{2,3}_{3,2}}$ & 12.509 & 1.231 \\ \cline{2-4}
\rule{0pt}{2.5ex}& $\mathcal{L}^{6,6}_{6(3,2),4(3,2)}\rightarrow \mathcal{L}_{7(3,2),2,1}^{5,4,3}$ & 12.655 & 1.246\\ \cline{2-4}
\rule{0pt}{2.5ex} &$\mathcal{K}_{7,3}\rightarrow \mathcal{L}^{4,4,4}_{4,3,2}$ & 12.663 & 1.247\\ \cline{2-4}
\rule{0pt}{2.5ex}& $\mathcal{K}_{5,3}\rightarrow \mathcal{L}_{9(3,2), 5}^{4,4}$ & 12.883 & 1.268 \\ \cline{2-4}
\rule{0pt}{2.5ex} & $\mathcal{K}_{7,2}\rightarrow \mathcal{K}_{7,2}$ & 13.696 &1.348 \\
\hline\hline
\multirow{7}{*}{$23$}\rule{0pt}{2.5ex} &$\mathcal{K}_{5,4}\rightarrow\mathcal{K}_{5,4}$ & 12.955 & 1.233\\  \cline{2-4}
\rule{0pt}{2.5ex}& $\mathcal{C}^{2,3}_{3,2},\mathcal{C}^{2,5}_{3,2},\mathcal{C}^{2,7}_{3,2}\rightarrow \boldsymbol{\mathcal{C}^{2,5}_{3,2}}$ & 12.959 & 1.234\\ \cline{2-4}
\rule{0pt}{2.5ex} & $\mathcal{L}^{5,5}_{9(3,2),4(3,2)}\rightarrow \mathcal{L}_{8(3,2),3,2}^{4,3,3}$& 12.962 & 1.234\\ \cline{2-4}
\rule{0pt}{2.5ex} &$\mathcal{K}_{7,3}\rightarrow\mathcal{K}_{7,3}$ & 13.162 & 1.253\\ \cline{2-4}
\rule{0pt}{2.5ex} & $\mathcal{K}_{5,3}\rightarrow \mathcal{L}_{11(5,2), 4}^{4,4}$ & 13.163 & 1.253\\ \cline{2-4}
\rule{0pt}{2.5ex} & $\mathcal{K}_{7,2}\rightarrow \mathcal{L}_{13(5,2), 2}^{4,4}$ & 13.171 & 1.254\\ \cline{2-4}
\hline\hline
\rule{0pt}{2.5ex}\multirow{6}{*}{$24$} & $\begin{matrix}\rule{0pt}{2.5ex}\mathcal{C}^{2,3}_{3,2},  \mathcal{C}^{2,5}_{3,2},\mathcal{C}^{2,7}_{3,2},\\ \rule{0pt}{2.5ex}\mathcal{K}_{5,4},\mathcal{L}^{5,5}_{9(3,2),5(3,2)},\\\rule{0pt}{2.5ex} \mathcal{L}^{6,6}_{6(3,2),6(3,2)}\end{matrix}\rightarrow \mathcal{L}^{6,6}_{6(3,2),6(3,2)}$ & 13.282 & 1.225 \\ \cline{2-4}
\rule{0pt}{2.5ex}
& $\mathcal{K}_{7,2}\rightarrow \mathcal{L}^{5,5}_{7(3,2),7(3,2)}$ & 13.284 &1.225 \\ \cline{2-4}
\rule{0pt}{2.5ex}
&$\mathcal{K}_{5,3}\rightarrow\mathcal{L}_{17(5,3),1}^{3,3}$ & 13.439 &1.239 \\ \cline{2-4}
\rule{0pt}{2.5ex}
&$\mathcal{K}_{7,3}\rightarrow\mathcal{L}_{8(3,2),4}^{6,6}$ & 13.628 & 1.257  \\
\hline\hline
\rule{0pt}{2.5ex}\multirow{5}{*}{$25$} & $\mathcal{C}^{2,3}_{3,2},\mathcal{C}^{2,5}_{3,2},\mathcal{C}^{2,7}_{3,2}\rightarrow \mathcal{L}^{6,6}_{7(3,2),6(3,2)}$ & 13.752 & 1.230 \\ \cline{2-4}
\rule{0pt}{2.5ex}
&$\mathcal{K}_{5,4}\rightarrow \mathcal{K}_{5,4}$ & 13.811&1.235 \\ \cline{2-4}
\rule{0pt}{2.5ex}
&$\mathcal{L}^{5,5}_{9(3,2),6(3,2)}\rightarrow \mathcal{L}_{15(4,3),2}^{4,4}$ & 13.861 & 1.240 \\ \cline{2-4}
\rule{0pt}{2.5ex}
&$\mathcal{K}_{5,3},\mathcal{K}_{7,3}\rightarrow \mathcal{K}_{7,3}$ &14.026  &1.255 \\ \cline{2-4}
\rule{0pt}{2.5ex}
&$\mathcal{K}_{7,2}\rightarrow \mathcal{L}_{11(5,2),4}^{5,5}$ & 14.112 & 1.262\\

\hline\hline
\rule{0pt}{2.5ex}\multirow{4}{*}{$26$} & $\mathcal{C}^{2,3}_{3,2},\mathcal{C}^{2,5}_{3,2},\mathcal{C}^{2,7}_{3,2}\rightarrow \mathcal{L}^{6,6}_{7(3,2),7(3,2)}$ & 14.057 &1.221 \\ \cline{2-4}
\rule{0pt}{2.5ex}
& $\mathcal{K}_{5,4}\rightarrow \mathcal{L}_{8(3,2),6(3,2)}^{6,6}$& 14.159 & 1.230\\ \cline{2-4}
\rule{0pt}{2.5ex}
&$\mathcal{K}_{5,3}, \mathcal{K}_{7,3}\rightarrow \mathcal{K}_{7,3}$ & 14.490 & 1.258 \\ \cline{2-4}
\rule{0pt}{2.5ex}
&$\mathcal{K}_{7,2}\rightarrow\mathcal{L}_{6,4,4}^{4,4,4}$& 14.697 &1.276  \\
\hline\hline
\rule{0pt}{2.5ex}\multirow{5}{*}{$27$} & $\mathcal{C}^{2,3}_{3,2},\mathcal{C}^{2,5}_{3,2},\mathcal{C}^{2,7}_{3,2}\rightarrow \boldsymbol{\mathcal{C}^{2,5}_{3,2}}$ &14.578 &1.231 \\ \cline{2-4}
\rule{0pt}{2.5ex}
& $\mathcal{K}_{7,2}\rightarrow\mathcal{L}_{9(3,2),6(3,2)}^{6,6}$ & 14.699 & 1.241 \\ \cline{2-4}
\rule{0pt}{2.5ex}
 &$\mathcal{L}^{5,5}_{11(3,2),6(3,2)}\rightarrow \mathcal{L}_{6(3,2),3,2}^{6,6,4}$  & 14.711&1.242 \\ \cline{2-4}
\rule{0pt}{2.5ex}
& $\mathcal{K}_{5,4}\rightarrow \mathcal{L}_{17(4,3),2}^{4,4}$& 14.743 &1.245 \\ \cline{2-4}
\rule{0pt}{2.5ex}
&$\mathcal{K}_{7,3}\rightarrow \mathcal{L}_{14(5,2),3}^{5,5}$ & 14.948 & 1.262\\ \cline{2-4}
\rule{0pt}{2.5ex}
 &$\mathcal{K}_{5,3}\rightarrow \mathcal{L}_{13(5,2),4}^{5,5}$ &14.977 &1.264 \\
\hline
\end{tabular}
\hspace*{-3cm}
\caption{Initial field configurations and the form of the numerical solution and then the energy $E$ and $E/Q^{3/4}$ for a comparison with Ward's conjectured bound for charges sixteen to twenty-seven. Solutions not formed of torus knots are marked in bold.}
\label{tab::energies1}
\end{table}

\begin{table}
\footnotesize\setlength{\tabcolsep}{2.5pt}
\hspace*{-3cm}
\centering
\begin{tabular}[t]{|c|c|c|c|}
\hline
$\phantom{Q}Q\phantom{Q}$ & Initial $\rightarrow$ final & $E$ & $E/Q^{3/4}$\\

\hline\hline
\rule{0pt}{2.5ex}\multirow{5}{*}{$28$} & $\mathcal{C}^{2,3}_{3,2}, \mathcal{C}^{2,5}_{3,2},\mathcal{C}^{2,7}_{3,2}, \mathcal{L}^{5,5}_{9(3,2),9(3,2)}\rightarrow\boldsymbol{\mathcal{C}^{2,3}_{3,2}}$ & 14.878 & 1.222\\ \cline{2-4}
\rule{0pt}{2.5ex}
&$\mathcal{K}_{5,4}\rightarrow\mathcal{L}_{4,2,2,2}^{6,4,4,4}$  & 15.023 & 1.234 \\ \cline{2-4}
\rule{0pt}{2.5ex}
&$\mathcal{K}_{5,3}\rightarrow\mathcal{L}_{6,3,3}^{6,5,5}$ & 15.198 & 1.249 \\ \cline{2-4}
\rule{0pt}{2.5ex}
&$\mathcal{K}_{7,2}\rightarrow \mathcal{K}_{8,3}$ & 15.382 &1.264 \\ \cline{2-4}
\rule{0pt}{2.5ex}
&$\mathcal{K}_{7,3}\rightarrow \mathcal{K}_{7,3}$ & 15.420 &1.267 \\ 
\hline\hline
\rule{0pt}{2.5ex}\multirow{6}{*}{$29$}& $\mathcal{C}^{2,3}_{3,2},\mathcal{C}^{2,5}_{3,2},\mathcal{C}^{2,7}_{3,2}\rightarrow \mathcal{L}^{6,6}_{9(3,2),8(3,2)}$ & 15.396 & 1.232\\ \cline{2-4}
\rule{0pt}{2.5ex}
 &$\mathcal{K}_{5,3}\rightarrow \mathcal{L}_{17(5,3),2}^{5,5}$ & 15.399 &1.232 \\ \cline{2-4}
\rule{0pt}{2.5ex}
& $\mathcal{K}_{5,4}\rightarrow \boldsymbol{\mathcal{C}^{2,7}_{3,2}}$ & 15.455 & 1.237\\ \cline{2-4}
\rule{0pt}{2.5ex}
&$\mathcal{L}^{5,5}_{11(3,2),8(3,2)}\rightarrow \mathcal{L}_{13(4,3),4}^{6,6}$ &15.478 & 1.239\\ \cline{2-4}
\rule{0pt}{2.5ex}
&$\mathcal{K}_{7,2}\rightarrow \mathcal{K}_{8,3} $& 15.852 &1.268 \\ \cline{2-4}
\rule{0pt}{2.5ex}
&$\mathcal{K}_{7,3}\rightarrow \mathcal{L}_{5,4,4}^{6,5,5}$ & 15.988 & 1.279 \\
\hline\hline
\rule{0pt}{2.5ex}\multirow{1}{*}{$30$} &$\mathcal{L}^{5,5}_{11(3,2),9(3,2)}\rightarrow \mathcal{L}_{18(5,3),2}^{5,5}$ & 15.713 & 1.226\\ \cline{2-4}
\rule{0pt}{2.5ex}
& $\mathcal{C}^{2,3}_{3,2},\mathcal{C}^{2,5}_{3,2},\mathcal{C}^{2,7}_{3,2}\rightarrow\boldsymbol{\mathcal{C}^{2,5}_{3,2}}$ & 15.802 &1.233  \\ \cline{2-4}
\rule{0pt}{2.5ex}
& $\mathcal{K}_{5,4}\rightarrow \boldsymbol{\mathcal{C}^{2,7}_{3,2}}$ & 15.851 & 1.237 \\
\hline\hline
\rule{0pt}{2.5ex}\multirow{1}{*}{$31$} &$\begin{matrix}\rule{0pt}{2.5ex} \mathcal{C}^{2,3}_{3,2},\mathcal{C}^{2,5}_{3,2},\mathcal{C}^{2,7}_{3,2}\\ \mathcal{L}^{5,5}_{13(3,2),8(3,2)} \end{matrix} \rightarrow \boldsymbol{\mathcal{C}^{2,7}_{3,2}}$ & 16.294 &1.240 \\ \cline{2-4}
\rule{0pt}{2.5ex}
& $\mathcal{K}_{5,4}\rightarrow \mathcal{L}_{15(4,3),4}^{6,6}$ & 16.319 & 1.242 \\
\hline
\end{tabular}
\hspace*{1ex}
\begin{tabular}[t]{|c|c|c|c|}
\hline
$\phantom{Q}Q\phantom{Q}$ & Initial $\rightarrow$ final & $E$ & $E/Q^{3/4}$\\
\hline\hline
\hline
\rule{0pt}{2.5ex}\multirow{3}{*}{$32$} &$\mathcal{L}^{5,5}_{11(3,2),11(3,2)}\rightarrow\mathcal{L}_{3,3,2,2}^{6,6,5,5}$ & 16.516 & 1.228 \\ \cline{2-4}
\rule{0pt}{2.5ex}
& $\mathcal{K}_{5,4}\rightarrow \mathcal{L}_{8(3,2),2,2,2}^{6,4,4,4}$ & 16.523 & 1.228 \\ \cline{2-4}
\rule{0pt}{2.5ex}
& $\mathcal{C}^{2,3}_{3,2},\mathcal{C}^{2,5}_{3,2},\mathcal{C}^{2,7}_{3,2}\rightarrow\boldsymbol{\mathcal{C}^{2,5}_{3,2} }$ & 16.736 & 1.244\\
\hline\hline
\rule{0pt}{2.5ex}\multirow{1}{*}{$33$} & $\mathcal{C}^{2,3}_{3,2},\mathcal{C}^{2,5}_{3,2}\rightarrow \mathcal{L}_{11(5,2),6(3,2)}^{8,8}$ & 16.925 & 1.229\\ \cline{2-4}
\rule{0pt}{2.5ex}
&$\mathcal{C}^{2,7}_{3,2},\mathcal{L}^{5,5}_{13(3,2),10(3,2)} \rightarrow \boldsymbol{\mathcal{H}_{3.609} }$ & 17.096 & 1.242\\ \cline{2-4}
\rule{0pt}{2.5ex}
&$\mathcal{K}_{5,4}\rightarrow \mathcal{L}_{3,3,2,1}^{7,7,5,5}$ & 17.149 & 1.246 \\
\hline\hline
\rule{0pt}{2.5ex}\multirow{1}{*}{$34$} & $\mathcal{K}_{5,4}\rightarrow \mathcal{L}_{4,4,2,2}^{6,6,5,5}$ & 17.334 & 1.231\\ \cline{2-4}
\rule{0pt}{2.5ex}
& $\mathcal{L}^{5,5}_{13(3,2),11(3,2)}\rightarrow \boldsymbol{\mathcal{H}_{3.609}}$ & 17.368 & 1.234\\ \cline{2-4}
\rule{0pt}{2.5ex}
& $\mathcal{C}^{2,3}_{3,2},\mathcal{C}^{2,5}_{3,2},\mathcal{C}^{2,7}_{3,2}\rightarrow\boldsymbol{\mathcal{C}^{2,3}_{3,2}}$ & 17.715 &1.258 \\
\hline\hline
\rule{0pt}{2.5ex}\multirow{3}{*}{$35$} & $\mathcal{K}_{5,4}\rightarrow\mathcal{L}_{13(4,3),6(3,2)}^{8,8}$ & 17.713 & 1.231\\ \cline{2-4}
\rule{0pt}{2.5ex} & $\mathcal{L}^{5,5}_{14(3,2),11(3,2)}\rightarrow\boldsymbol{\mathcal{H}_{3.609}}$ & 17.903 & 1.244\\ \cline{2-4}
\rule{0pt}{2.5ex}
& $\mathcal{C}^{2,3}_{3,2},\mathcal{C}^{2,5}_{3,2},\mathcal{C}^{2,7}_{3,2} \rightarrow \mathcal{L}^{7,7}_{11(3,2),10(3,2)}$ & 18.331& 1.274\\
\hline\hline
\rule{0pt}{2.5ex}\multirow{2}{*}{$36$} & $\mathcal{C}^{2,3}_{3,2},\mathcal{C}^{2,7}_{3,2}\rightarrow\boldsymbol{\mathcal{L}_{26(2,3;3,2),2}^{4,4}}$ & 18.019 & 1.226\\ \cline{2-4}
\rule{0pt}{2.5ex}
&$\mathcal{C}^{2,5}_{3,2}\rightarrow \boldsymbol{\mathcal{H}_{3.609}}$ & 18.178 & 1.237\\ \cline{2-4}
\rule{0pt}{2.5ex}
& $\mathcal{K}_{5,4}\rightarrow \mathcal{L}_{15(4,3),2,1}^{8,5,5}$ & 18.275 & 1.243 \\ \cline{2-4}
\rule{0pt}{2.5ex}
&$\mathcal{L}^{5,5}_{13(3,2),13(3,2)}\rightarrow \boldsymbol{\mathcal{L}_{20\mathcal{H}_{2.828},4 }^{6,6}}$ & 18.400 & 1.252\\
\hline
\end{tabular}
\hspace*{-3cm}
\caption{Initial field configurations and the form of the numerical solution and then the energy $E$ and $E/Q^{3/4}$ for a comparison with Ward's conjectured bound for charges twenty-eight to thirty-six. Solutions not formed of torus knots are marked in bold.}
\label{tab::energies2}
\end{table}

For charges up to twenty-one we find that initial conditions which take the form of cable knots are unstable and result in fields resembling torus knots or links comprised of torus knots with unknotted components. For this range of charge cable knot solutions are not expected since the charge due to the winding of the field is small - or even negative. However they give a complicated unstable initial field configurations and so provide a good starting condition to find energy minima. For each charge we find that there are a range of link solutions comprising of a trefoil and an approximately axial hopfion with the charge shared between components in various ways. At charge sixteen we recover the solution $16\mathcal{L}_{8(3,2),2}^{2,2}$ first found in \cite{sutcliffe2007} (where it was denoted $\mathcal{X}_{16}$), and also find lower energy configurations $16\mathcal{K}_{5,3}$ and $16\mathcal{L}_{4,4}^{4,4}$. At charge seventeen we find a solution of the form $17\mathcal{L}_{2,1,1,1}^{3,3,3,3}$ which is the first known example of a four-component link in the model, however the torus link $17\mathcal{L}_{8(3,2),3}^{3,3}$ was the lowest-energy configuration found. For charges eighteen and nineteen torus knot configurations were found to have the lowest energy, $18\mathcal{K}_{5,3}$ and $19\mathcal{K}_{5,4}$ respectively. The lowest energy configuration found for charge twenty is the link $20\mathcal{L}_{3,3,2}^{4,4,4}$, and for charge twenty-one the link $21\mathcal{L}_{13(4,3),2}^{3,3}$ has the lowest energy.

At charge twenty-two we find the first example of hopfion with the field taking the form of a cable knot. We find that there is a local energy minimum of the form $22\mathcal{C}^{2,3}_{3,2}$ which exceeds the lowest energy solution found for that charge, a solution of the form $22\mathcal{L}_{8(3,2),2,2}^{4,3,3}$, by less than $1\%$. Similarly for charge twenty-three we again find a cable knot, this time it has the form of $23\mathcal{C}^{2,5}_{3,2}$ and is even closer to the lowest energy solution found, $23\mathcal{K}_{5,4}$. We find that for charges twenty-four to twenty-six the cable-knotted fields relax to resemble cable links, which are the minimal energy solutions found for these charges. These are the first examples known of solutions which are of the form of a link with two components being torus knots. We then find that cable knots occur again for $27\mathcal{C}^{2,5}_{3,2}$, $28C^{2,3}_{3,2}$ and that this time they appear as the energy minimum found. For charge twenty-nine we again find a cable link being the minimal energy configuration, before finding local energy minima cable knots $30\mathcal{C}^{2,5}_{3,2}$, $31\mathcal{C}^{2,7}_{3,2}$, $32\mathcal{C}^{2,5}_{3,2}$ and $34\mathcal{C}^{2,3}_{3,2}$. When one should expect cable knotted fields to appear in the theory seems to be a complicated question, with many possible factors influencing this such as the twist of the field, the length and the self-interaction of strands. We note that for the cases of cable knots with odd charge that we can see that the centre of solutions appears longer than those with even charges. This can be understood since it is energetically favourable for the twisting about the core to align with a certain phase - as in the baby Skyrme model. If we approximate the twist per unit length as approximately constant, this can happen more smoothly for an odd number of twists, and hence an even charge. When this becomes unfavourable we see a transition to cable links.

For some charges we find that the cable knotted initial fields relax to other structures. For charge thirty-six the lowest energy solution we have found is a link where one component is of the form of a charge twenty-six cable knot $\mathcal{C}^{2,3}_{3,2}$. We extend our notation for links to denote this $36\mathcal{L}_{26(2,3;3,2),2}^{4,4}$.  For charge eighteen we find a local minimum of the energy which is the first known solution with the structure of a hyperbolic knot. A knot is said to be hyperbolic if the complement of the knot has a metric with constant negative curvature \cite{riley1975,adams2004}. One can then calculate the volume of this complement with resepect to this metric to find the hyperbolic volume. In this case one can use the SnapPy program \cite{snappy} to calculate that it has hyperbolic volume 4.85117 and can identify it as knot $10_{139}$ in Rolfsen notation, or $k5_{22}$ in Callahan-Dean-Weeks-Champanerkar-Kofman-Patterson notation. We denote this by its hyperbolic volume as $\mathcal{H}_{4.851}$. For charges thirty-four to thirty-six we find further examples of knotted structures which take the form of hyperbolic knots. We see that for these charges the fields are knotted in the same way, and one can calculate that it has hyperbolic volume 3.60869 and can classify this knot as the knot $k4_4$. We therefore denote it as $\mathcal{H}_{3.609}$. Finally, we also find at charge thirty-six a link with one of the components being a hyperbolic knot. In this case we can identify it as the knot $k3_1$, which has hyperbolic volume 2.82812 and so denote the link as $\mathcal{L}_{20\mathcal{H}_{2.828},4 }^{6,6}$


\section{Conclusion}
In this paper we have presented new rational maps which generate fields with a knotted structure of the form of cable knots. We have then also demonstrated that fields with such configurations are stable solutions in the Skyrme-Faddeev model and in doing so have found the first known hopfions which are not of the form of torus knots. We have also found examples of the first known hyperbolic knots in the model.

There are a number of interesting phenomenon which it would be interesting to develop further. There are a number of different knotted structures which would be interesting to explore in the model. In this current study hopfions taking the form of hyperbolic knots have arisen from unstable ansatz of different structures. A rational map which generates a field of the form of hyperbolic knots would enable the study of these knots in more detail. Cable knots themselves have a natural extension, iterated torus knots, created by embedding these cable knots themselves onto a torus knot. It would be intriguing to see whether this iterated behaviour is something we can expect to occur in this model for large charge.

Similar knotted structures are also evident in a range of related models, such as the Nicole \cite{nicole1978,gillard2010}, Aratyn-Ferreira-Zimerman \cite{aratyn1999,gillard2011} and extended Skyrme-Faddeev \cite{gladikowski1997} models. One would expect similarly knotted fields to appear as solutions in these models and it would be interesting to check whether these occur. The construction of knotted fields is used in a range of models in an analogous way to the Skyrme-Faddeev model, such as in knotted light fields \cite{dennis2010}, and so our construction might have applications in these situations.

\section*{Acknowledgements}
The author would like to thank Daniel Jones, Andrew Lobb, Martin Speight and Paul Sutcliffe for helpful discussions and suggestions. Work was undertaken while funded by an STFC studentship.

\bibliographystyle{prsty}
\bibliography{cable}

\begin{thebibliography}{10}

\bibitem{faddeev1975}
L. Faddeev, Preprint {IAS} Print-75-{QS}70 (Inst. Advanced Study, Princeton,
  {NJ}, 1975)  .

\bibitem{faddeev1997}
L. Faddeev and A.~J. Niemi, Nature {\bf 387},  58  (1997).

\bibitem{skyrme1961}
T.~H.~R. Skyrme, Proc. R. Soc. Lond. A {\bf 260},  127  (1961).

\bibitem{piette1995}
B.~M. A.~G. Piette, B.~J. Schroers, and W.~J. Zakrzewski, Z. Phys. C {\bf 65},
  165  (1995).

\bibitem{faddeev1998}
L. Faddeev and A.~J. Niemi, Phys. Rev. Lett. {\bf 82},  1624  (1999).

\bibitem{kawaguchi2008}
Y. Kawaguchi, M. Nitta, and M. Ueda, Phys. Rev. Lett. {\bf 100},  180403
  (2008).

\bibitem{gladikowski1997}
J. Gladikowski and M. Hellmund, Phys. Rev. D {\bf 56},  5194  (1997).

\bibitem{battye1998}
R.~A. Battye and P.~M. Sutcliffe, Phys. Rev. Lett. {\bf 81},  4798  (1998).

\bibitem{battye1999}
R.~A. Battye and P.~M. Sutcliffe, Proc. R. Soc. Lond. A {\bf 455},  4305
  (1999).

\bibitem{hietarinta1999}
J. Hietarinta and P. Salo, Phys. Lett. B {\bf 451},  60  (1999).

\bibitem{hietarinta2000}
J. Hietarinta and P. Salo, Phys. Rev. D {\bf 62},  081701  (2000).

\bibitem{ward2000}
R. Ward, Phys. Lett. B {\bf 473},  291   (2000).

\bibitem{sutcliffe2007}
P. Sutcliffe, Proc. R. Soc. A {\bf 463},  3001  (2007).

\bibitem{thurston1986}
W.~P. Thurston, Ann. Math. {\bf 124},  pp. 203  (1986).

\bibitem{adams2004}
C. Adams, {\em The Knot Book} (AMS, Providence, R.I., 2004).

\bibitem{bott1995}
R. Bott and L.~W. Tu, {\em Differential Forms in Algebraic Topology} (Springer,
  New York, 1995).

\bibitem{vakulenko1979}
A.~F. {Vakulenko} and L.~V. {Kapitanskii}, Sov. Phys. Dokl. {\bf 24},  433
  (1979).

\bibitem{kundu1982}
A. Kundu and Y.~P. Rybakov, J. Phys. A {\bf 15},  269  (1982).

\bibitem{ward1999}
R.~S. Ward, Nonlinearity {\bf 12},  241  (1999).

\bibitem{brieskorn1986}
E. Brieskorn and H. Kn\"{o}rrer, {\em Plane Algebraic Curves} (Birkhäuser
  Basel, Basel; Boston, 1986).

\bibitem{eisenbud1985}
D. Eisenbud and W.~D. Neumann, {\em Three-dimensional Link Theory and
  Invariants of Plane Curve Singularities} (PUP, Princeton, 1985).

\bibitem{kahler1929}
E. Kahler, Math. Zeits. {\bf 30},  188  (1929).

\bibitem{wall2004}
C.~T.~C. Wall, {\em Singular Points of Plane Curves} (CUP, Cambridge, 2004).

\bibitem{riley1975}
R. Riley, Mathematika {\bf 22},  141  (1975).

\bibitem{snappy}
M. Culler, N.~M. Dunfield, and J.~R. Weeks, Snap{P}y, a computer program for
  studying the topology of $3$-manifolds, Available at
  \url{http://snappy.computop.org}.

\bibitem{nicole1978}
D. Nicole, J. Phys. {\bf G4},  1363  (1978).

\bibitem{gillard2010}
M. Gillard and P. Sutcliffe, J. Math. Phys. {\bf 51},  122305  (2010).

\bibitem{aratyn1999}
H. Aratyn, L. Ferreira, and A. Zimerman, Phys. Lett. B {\bf 456},  162  (1999).

\bibitem{gillard2011}
M. Gillard, Nonlinearity {\bf 24},  2729  (2011).

\bibitem{dennis2010}
M.~R. Dennis {\it et~al.}, Nat Phys {\bf 6},  118  (2010).

\end{thebibliography}
\end{document}